\documentclass[10pt,letterpaper]{article}
\usepackage[top=0.85in,left=2.75in,footskip=0.75in,marginparwidth=2in]{geometry}

\usepackage[utf8]{inputenc}

\usepackage{cite}

\usepackage{nameref,hyperref}
\usepackage{amsmath,amssymb} 
\usepackage{capt-of}
\usepackage{graphicx,times}
\usepackage{float}
\usepackage{color}
\usepackage{array}
\usepackage{multirow}
\usepackage{multicol}
\usepackage{fixltx2e}
\usepackage{verbatim}

\usepackage[right]{lineno}

\usepackage{microtype}
\DisableLigatures[f]{encoding = *, family = * }

\raggedright
\usepackage{changepage}

\usepackage[aboveskip=1pt,labelfont=bf,labelsep=period,singlelinecheck=off]{caption}

\makeatletter
\renewcommand{\@biblabel}[1]{\quad#1.}
\makeatother

\usepackage{lastpage,fancyhdr,graphicx}
\usepackage{epstopdf}
\fancyheadoffset[L]{2.25in}
\fancyfootoffset[L]{2.25in}

\usepackage{color}

\definecolor{Gray}{gray}{.25}

\usepackage{graphicx}

\usepackage{sidecap}

\usepackage{wrapfig}
\usepackage[pscoord]{eso-pic}
\usepackage[fulladjust]{marginnote}
\reversemarginpar

\begin{document}
\vspace*{0.35in}

\begin{flushleft}
{\Large
\textbf\newline{A Deep Learning based Approach to Reduced Order Modeling for Turbulent Flow Control using LSTM Neural Networks}
}
\newline
\\
Arvind T. Mohan\textsuperscript{*} \&
Datta V. Gaitonde\textsuperscript{}
\\
\bigskip
\bf{} Mechanical and Aerospace Engineering, The Ohio State University, Columbus, OH
\\
\bigskip
* \textit{Currently at Center for Nonlinear Studies, Los Alamos National Laboratory (arvindm@lanl.gov)}

\end{flushleft}

\section*{Abstract}
Reduced Order Modeling (ROM) for engineering applications has been a major research focus in the past few decades due to the  unprecedented physical insight into turbulence offered by high-fidelity CFD. The primary goal of a ROM is to model the key physics/features of a flow-field without computing the full Navier-Stokes (NS) equations. This is accomplished by projecting the high-dimensional dynamics to a low-dimensional subspace, typically utilizing dimensionality reduction techniques like Proper Orthogonal Decomposition (POD), coupled with Galerkin projection. In this work, we demonstrate a deep learning based approach to build a ROM using the POD basis of canonical DNS datasets, for turbulent flow control applications. We find that a type of Recurrent Neural Network, the Long Short Term Memory (LSTM) which has been primarily utilized for problems like speech modeling and language translation, shows attractive potential in modeling temporal dynamics of turbulence. Additionally, we introduce the Hurst Exponent as a tool to study LSTM behavior for non-stationary data, and uncover useful characteristics that may aid ROM development for a variety of applications.

\section{Introduction}
Turbulence in fluids has been an active research topic due to its impact on a wide variety of applications, including those in aeronautics, transportation, energy generation systems and weather forecasting. Several experimental studies of turbulent flows in various canonical and practical cases have enhanced the understanding of turbulent behavior, leveraging it to design more efficient systems. Although experiments have been invaluable, complementary computational fluid dynamics (CFD) efforts have the ability to gain a more detailed insight into the physics of turbulent flows, especially in situations where experimentation had been too expensive and/or impractical. With the availability of increased computing power in the recent years, high-fidelity CFD techniques like Large Eddy Simulation (LES) and Direct Numerical Simulation (DNS) have made it feasible to study turbulence with an unprecedented level of detail. However, with an increase in fidelity there arise some significant challenges and opportunities. LES/DNS computations are very time intensive and expensive, therefore they are primarily used as a research tool to study the fundamental physics of turbulence. Additionally, these computations generate extremely high dimensional large datasets typically contain millions of degrees of freedom, which are often intractable to efficiently handle and analyze.

As a result, techniques for modeling the high fidelity dynamics of turbulence, while significantly minimizing the steep computation and data storage costs associated with LES/DNS have been a subject of active research~\cite{rowley2004model}. Such efforts to model spatio-temporal dynamics of turbulence in a low dimensional space are generally referred to as Reduced Order Models (ROMs). ROMs have two primary objectives: a) The ability to model the key dynamics/coherent features of the turbulent flow, and b) Provide an efficient means of data compression for LES/DNS datasets. A major application of such ROMs is to design flow control systems for turbulence, where their low computational cost and efficient models make them ideal candidates for building control logic for actuators~\cite{ito1998reduced,ravindran2000reduced}. Indeed, this capability of ROMs has been much sought after and has been a subject of considerable research, due to its wide-ranging applications in aerospace and mechanical engineering.  

Various methods of building ROMs exist, the common underlying theme being extracting the key features in the flow-field, preferably from a high fidelity experimental or computational data source. These extracted features are carefully chosen such that they represent dominant spatio-temporal dynamics computed by the Navier-Stokes equations. Mathematically, the goal here is \textit{model reduction} i.e. representing high dimensional data in a low-dimensional subspace, which is essential to reduce computational and data handling costs~\cite{noack2011reduced}. In literature, various types of ROMs have been demonstrated for canonical problems. Noack~\cite{noack2011reduced} showed ROM based control for fluid flows. Kaiser~\cite{kaiser2014cluster} used cluster based modeling to extract features and build adaptive ROMs. Duriez et. al.~\cite{duriez2017machine} used ROM to build a Machine learning based control for various nonlinear dynamical systems.  Generally, the most commonly adopted model reduction technique is the Proper Orthogonal Decomposition (POD)~\cite{berkooz1993proper}, as POD reduced basis (or modes) are mathematically optimal~\cite{holmes2012turbulence} for any given dataset. After model reduction, the next step is to use the reduced basis for modeling the flow at \textit{future} time instants. A highly popular technique is the Galerkin projection (GP) approach, which has been documented extensively in literature~\cite{burkardt2006pod, carlberg2011efficient, rapun2010reduced}. The crux of the GP method lies in the use of spatio-temporal dynamics captured by the reduced basis (such as POD) which can then be evolved them in time, instead of the full Navier-Stokes equations. The use of the reduced basis with ODEs ensures that the computation is much cheaper, since they contain considerably fewer degrees of freedom. On the other hand, Navier-Stokes simulations utilize PDEs and considerably large mesh sizes which drastically increase computational costs. Therefore, GP based approaches have seen widespread popularity and have been demonstrated for several canonical problems~\cite{rowley2004model,rowley2006dynamics}. However, GP models do not typically account for spatial variations in the flow and are known to become unstable under different conditions, even for canonical cases~\cite{akhtar2009stability,iollo2000stability,barone2009stable,rempfer2000low}. As such, considerable research efforts are being devoted to improving the stability of GP models, with some efforts focusing on Galerkin-free formulations~\cite{shinde2016galerkin}.

As a potential alternative, this paper explores a non-Galerkin projection based approach to ROM using deep learning~\cite{wang2017model}. Deep learning is a subset of machine learning which refers to the use of highly multilayered neural networks to understand (or ``learn") a complex dataset with the intention of predicting some features of that dataset~\cite{lecun2015deep}. The neural network does so by internally extracting a set of patterns or associations between different variables/parameters in the dataset, and subsequently using these patterns to predict a variable of interest (the output) given some input variable(s)~\cite{bengio2009learning,ivakhnenko1971deeplearning}. In recent years, deep learning has shown considerable promise in modeling complex datasets in diverse fields like imaging, finance, robotics etc. Application of machine learning (and deep learning) in fluid mechanics is an emerging area of research, with several efforts by Duraisamy~\cite{duraisamy2015new}, Wang~\cite{wang2016physics} and Ling~\cite{ling2015evaluation} focusing on improving the accuracy of CFD simulations using data-driven approaches. Apart from CFD, reduced order modeling appears to be an ideal candidate to be approached using machine learning algorithms, due to the sheer data-driven nature of the problem. The focus of this work is to exploit neural networks to ``learn" the key dynamics of turbulent flows from high-fidelity simulation databases and use them to generate ROMs for flow control applications. These ROMs can then be employed to model the flow-field at future time instants. 

Different types of neural networks (NNs) exist for various types of data, and a choice must be made based on the application. For instance, Convolutional neural networks (CNNs) are well suited for modeling image/spatial data fields, whereas Recurrent neural networks (RNNs) are specialized NNs for modeling sequential data, like time signals. There are a few variants of RNNs, with the most popular variant being the Long Short Term Memory (LSTM) Network, as it explicitly accounts for the \textit{memory} in a signal, without the vanishing gradient problem~\cite{hochreiter1998vanishing} often seen in RNNs. In this context, \textit{memory} intuitively refers to the pattern seen in several man-made and natural phenomena, where the current state of a variable may be significantly correlated to one or more of its previous states, implying that the phenomenon ``remembers" its past. Data obtained from turbulent flows, like other highly nonlinear dynamical systems, may exhibit memory effects which have to be accommodated for accurate predictive models. Therefore, this preliminary work explores the feasibility of LSTMs to build reduced order models for complex systems like turbulence.

\section{Long Short-Term Memory Neural Networks}

The aforementioned goal of predicting the evolution of a flow-field through its POD temporal coefficients is a sequence modeling problem in machine learning. Sequence prediction is different from other types of learning problems, since it imposes an order on the observations that must be preserved when training models and making predictions. Recurrent Neural Networks (RNNs) are a type of neural network specially designed for such problems, where the addition of sequence is a new dimension to the function being approximated. 

The Long Short-Term Memory (LSTM), network is a special variant of RNN, which overcomes stability bottlenecks encountered in traditional RNNs (like the Vanishing Gradient), enabling its practical application. LSTMs can also learn and harness the temporal dependence from the data. Additionally, LSTMs also utilize their internal memory; such that the predictions are conditional on the recent context in the input sequence, not what has just been presented as the current input to the network. For instance, they can be shown one observation at a time from a sequence and can learn what observations it has seen previously are relevant, and how they can be used to making a prediction. 

\begin{figure}
	\centering
	\fbox{\includegraphics[width=0.65\linewidth]{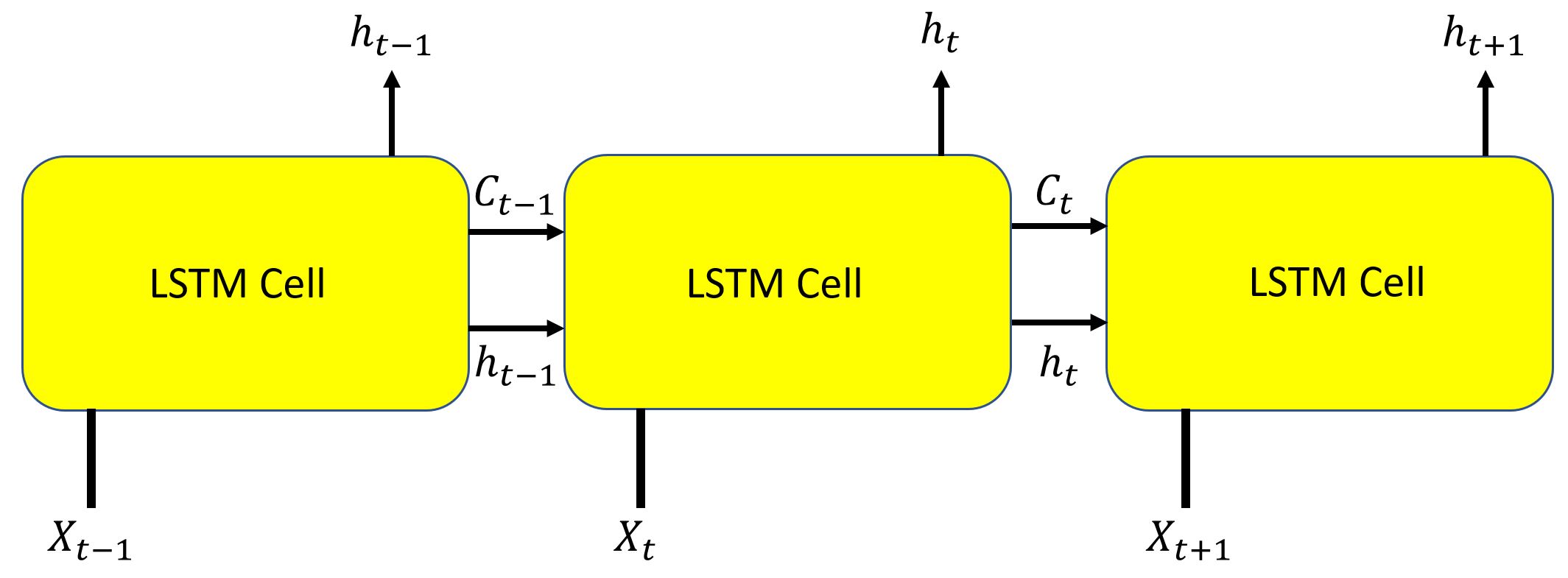}}
	\caption{LSTM Layout with Cell Connections}
	\label{lstmchain}
\end{figure}

The basic architecture of the LSTM NN is now outlined. The LSTM networks are different from other deep learning architectures like convolutional neural networks (CNNs), in that the typical LSTM cell contains three \textit{gates}: The \textbf{input} gate, \textbf{output} gate and the \textbf{forget} gate. The LSTM regulates the flow of training information through these gates by selectively adding information (input gate), removing (forget gate) or letting it through to the next cell (output gate). A schematic of the cells connected in a recurrent form is shown in Fig.~\ref{lstmchain}.

\begin{figure}
	\centering
	\fbox{\includegraphics[width=0.65\linewidth]{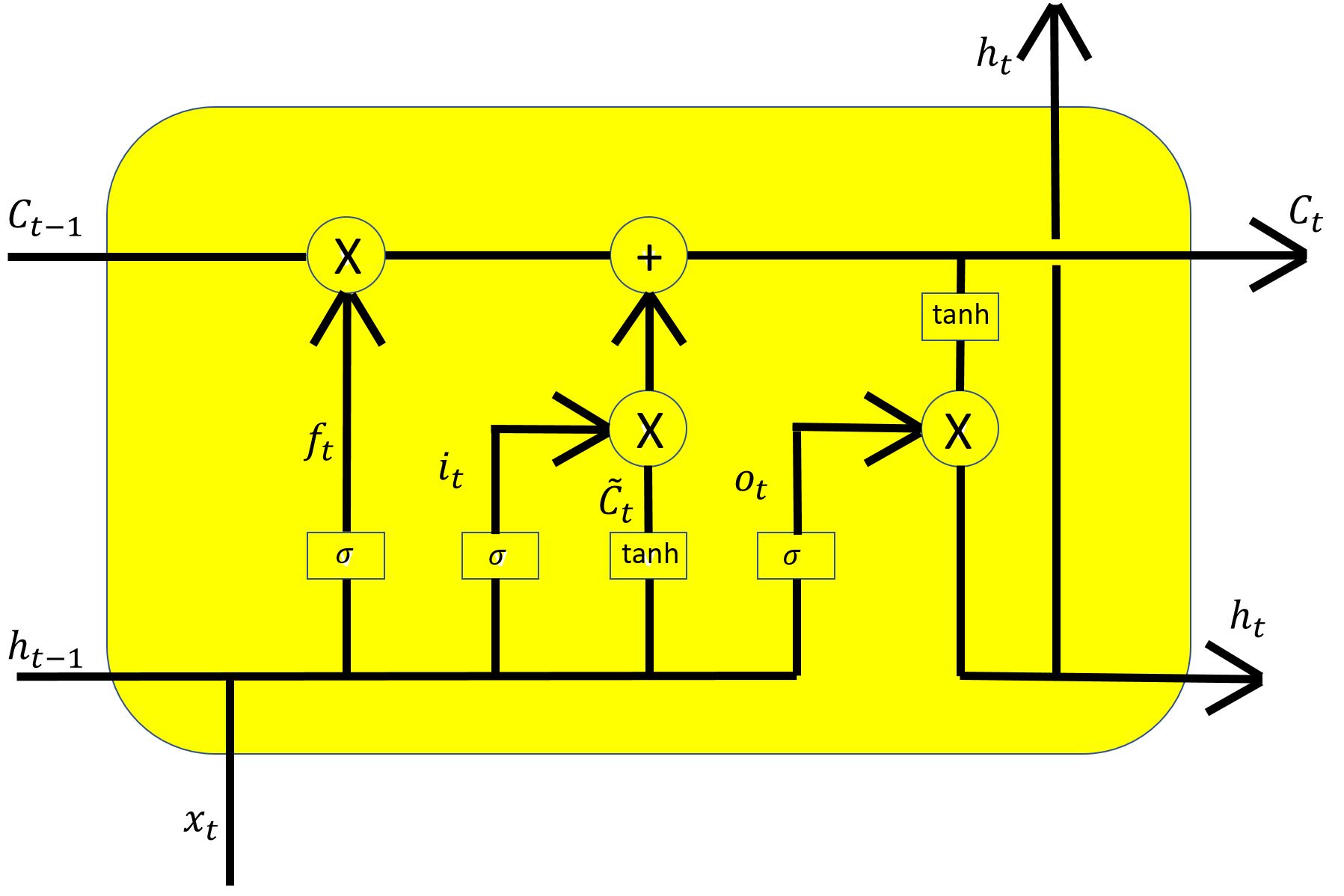}}
	\caption{Architecture of a LSTM Cell with Various Gates}
	\label{lstmcell}
\end{figure}

The input gate is represented by $i$, output gate by $o$ and forget gate by $f$. The cell state is represented as $C$ and the cell output is given by $h$, while the cell input is denoted as $x$. Consider the structure of a LSTM cell in Fig.~\ref{lstmcell}. The equations to compute its gates and states are as follows,

\begin{align}
    f_{t}\,&=\,\sigma \left(W_{f} \cdot \left[h_{t-1},x_{t}\right] + b_{f} \right)  \\
    i_{t}\,&=\,\sigma \left(W_{i} \cdot \left[h_{t-1},x_{t}\right] + b_{i} \right) \\
    \Tilde{C}_{t}\,&=\, tanh\left(W_{C} \cdot \left[h_{t-1},x_{t}\right] + b_{C} \right) \\    
    C_{t} \,&=\, f_{t}*C_{t-1} + i_{t}*\Tilde{C}_{t} \\
    o_{t}\,&=\,\sigma \left(W_{o} \cdot \left[h_{t-1},x_{t}\right] + b_{o} \right) \\    
    h_{t} \,&=\, o_{t}*tanh \left(C_{t}\right)
\end{align}
$W$ are the weights for each of the gates and $\Tilde{C}$ is the updated cell state. These states are propagated ahead through the network, as shown in Fig.~\ref{lstmchain} and weights are updated by backpropagation through time. The forget gate plays a crucial role in reducing over-fitting by not retaining all information from the previous time steps. This arrangement of gates and selective information control is also the key reason why LSTMs do not suffer from the vanishing gradient problem which plagued traditional RNNs. As a result, LSTMs are a powerful tool to model non-stationary datasets. A more detailed and informal introduction to LSTM can be found in Ref.~\citen{colahLSTM}.

In this work, we explore two variants of LSTMs: a) The \textbf{traditional LSTM} algorithm by Hochreiter~\cite{hochreiter1997long}, and b) The \textbf{Bidirectional LSTM} by Graves et al.~\cite{graves2005framewise,graves2005bidirectional}.
\begin{figure}
	\centering
	\fbox{\includegraphics[width=0.7\linewidth]{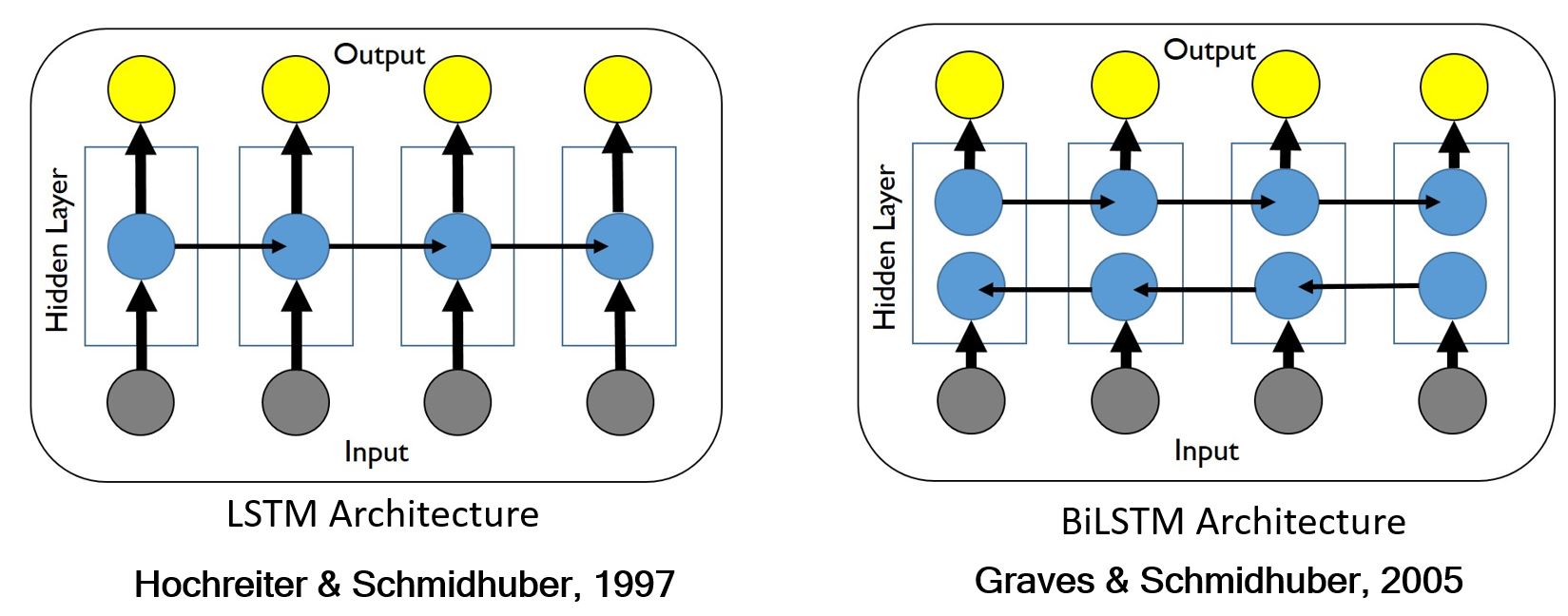}}
	\caption{LSTM and BiLSTM Architectures}
	\label{lstmcomparison}
\end{figure}
Figure~\ref{lstmcomparison} shows the general schematic of a LSTM NN and Bidirectional LSTM (BiLSTM). Each hidden layer cell in the LSTM has an input which is dependent on the computation of the cell handling the data at the previous time instant. This explicit accommodation of memory in a sequence makes LSTM a very useful approach for sequential modeling. Further information can be found in Schmidhuber~\cite{schmidhuber2015deep}. In contrast, BiLSTM has a two way flow of information. In this network, the sequence is trained by two LSTM networks, one in forward and the other in reverse direction. Both these networks are connected to the same output layer, and this architecture has shown improved accuracy for  language modeling~\cite{huang2015bidirectional,marchi2014multi,fan2014tts} and related applications. The reader is referred to Graves~\cite{graves2005framewise,graves2005bidirectional} for further information.
Therefore, in addition to exploring the feasibility of LSTMs for turbulence ROMs, an ancillary goal is to determine if BiLSTM offers improved performance compared to LSTMs.

\section{Methodology}

Neural networks excel at ``learning" data and using it for predictive modeling. One of the primary requirements for employing deep learning techniques is the availability of sufficient \textit{training data}. Lack of training data can jeopardize
the accuracy of the predictive model, even if state-of-the-art NN architectures and computing power are available. Therefore, any predictive modeling endeavor for nonlinear dynamical systems such as turbulence
should ensure access to high quality training data, as a first step.
In this work, we use two DNS databases from the Johns Hopkins turbulence database (JHTB)~\cite{graham2016web} to demonstrate the LSTM-ROM (LSTM-ROM), as it is a well known canonical case~\cite{patterson1971spectral}: a) \textit{Forced Isotropic Turbulence}, b) \textit{Magnetohydrodynamic Turbulence}.

The \textit{Forced Isotropic Turbulence} dataset (ISO) is sourced from 3-D DNS Navier Stokes simulations solved using a spectral method on a grid size of $1024^{3}$, with $5023$ timesteps of 3-D data frames saved for analysis. This corresponds to $10$ seconds of flow data, which makes this an extremely high-fidelity dataset. Similarly, The \textit{Magnetohydrodynamic Turbulence} dataset (MHD) is from a 3-D DNS Navier Stokes simulation also solved on a grid size of $1024^{3}$, with $1024$ timesteps of 3-D data frames saved for analysis. We use the U-velocity field for all results shown in this work. This corresponds to $2.056$ seconds of high-fidelity flow data. However, due to the high computational cost of a 3-D DNS, the JHTB provides these datasets only at a single Reynolds number, whereas a deep learning approach would ideally require multiple closely related datasets.

\begin{figure}
	\centering
	\fbox{\includegraphics[width=0.8\linewidth]{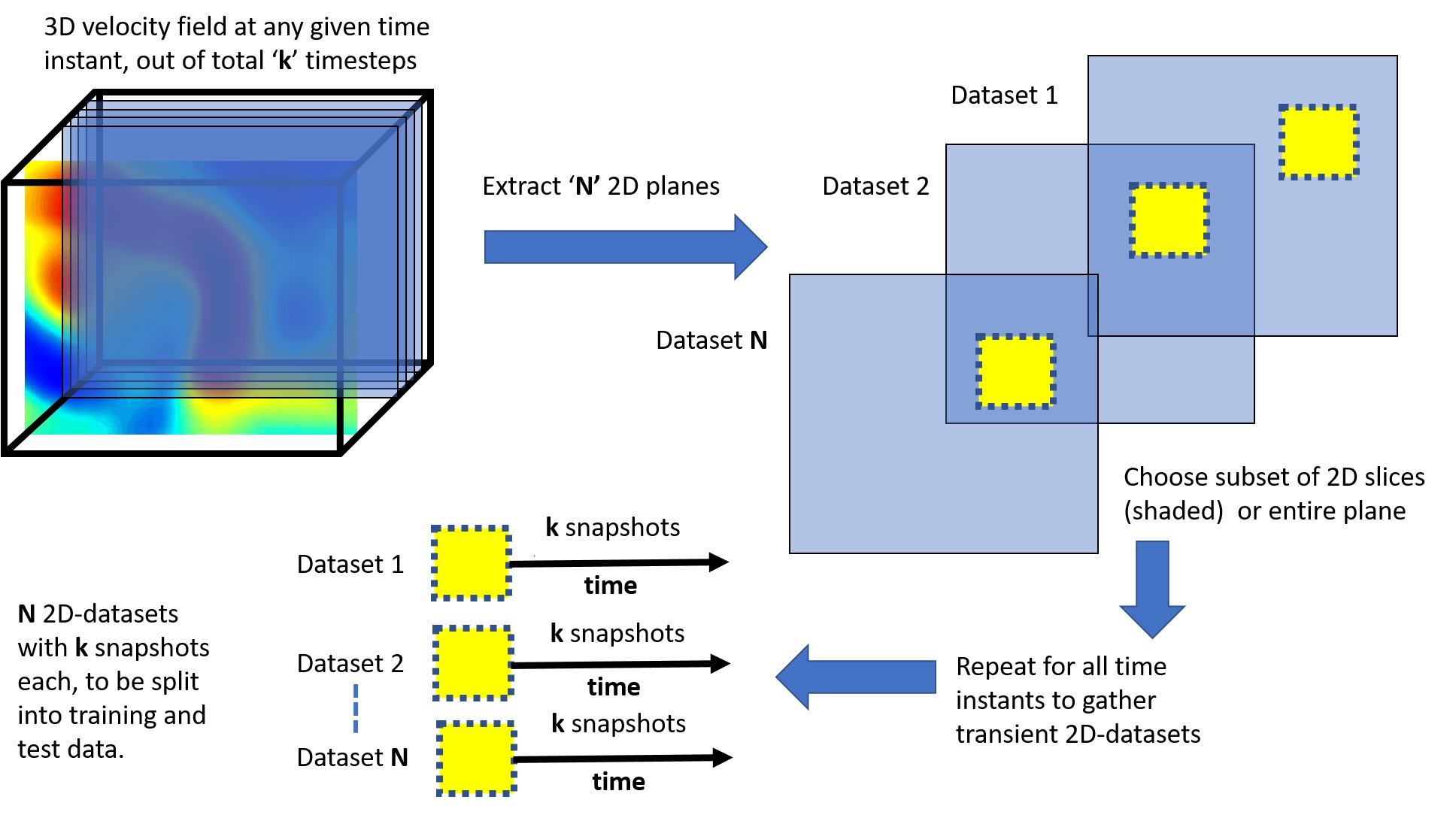}}
	\caption{Extracting 2-D datasets from a 3-D flow field}
	\label{dataExtract}
\end{figure}

\begin{figure}
	\centering
	\fbox{\includegraphics[width=1\linewidth]{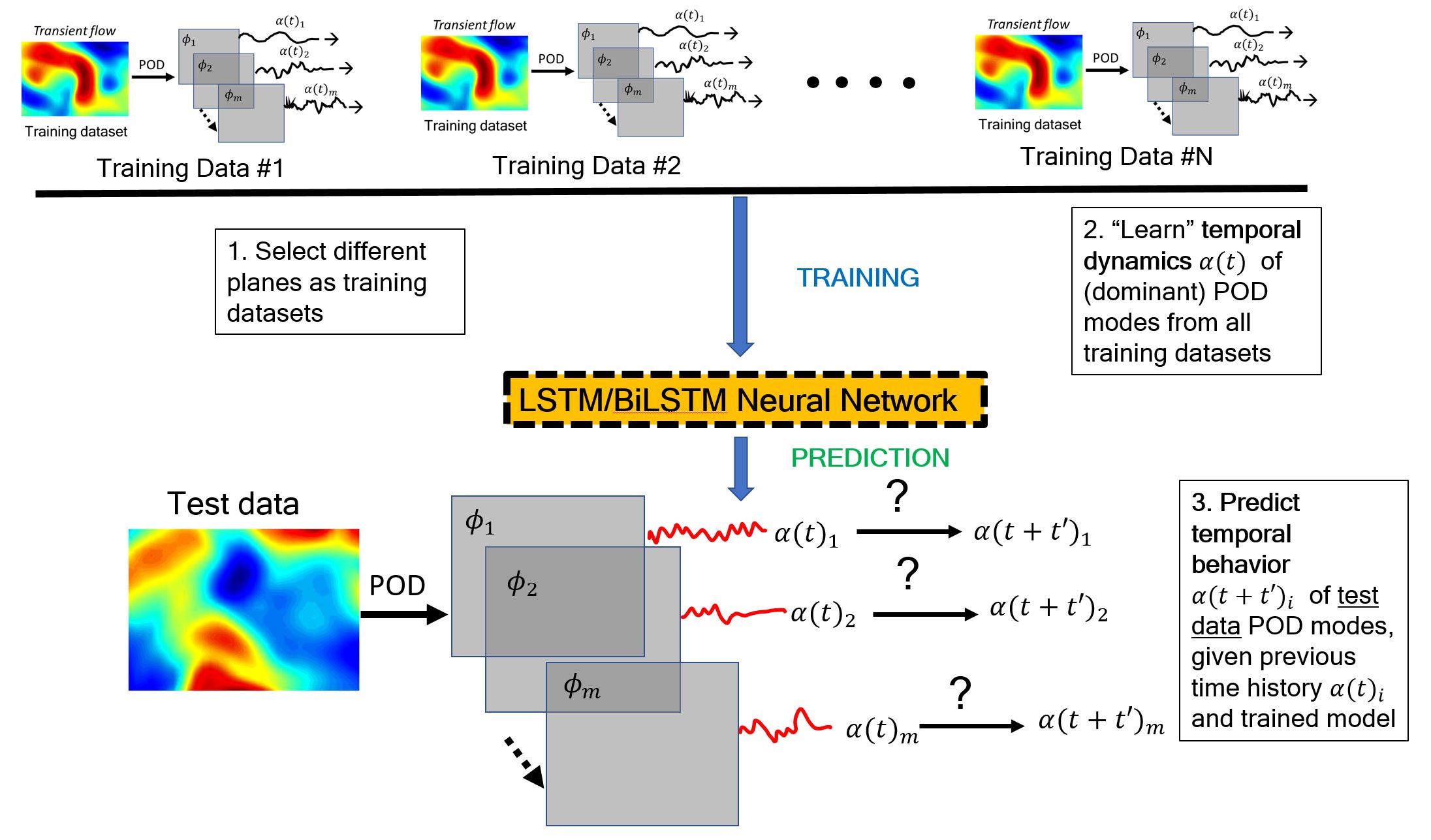}}
	\caption{LSTM-ROM Methodology}
	\label{lstmrom}
\end{figure}

In order to circumvent this  limitation, the 3-D ISO dataset is decomposed into several 2-D planes of the same size. In doing so,
a large number of unique training ``datasets"  are created. The process is explained in the Fig.~\ref{dataExtract} schematic, where ``N" different planes are extracted, while the total number of snapshots (k=5023 for ISO and k=1024 for MHD) for each 2D dataset (or a smaller subset) is retained. Additionally, as the flows are homogeneous isotropic turbulence and MHD, it is seen that several planes in the same region of the flow have some qualitative similarity, though several differences exist which are non-trivial. This is important, since any DL model needs training data which has strong similarity to the dynamics that have to be modeled. A major advantage of well designed NN models is that they account for the dissimilarities in various training datasets, and extract only the key patterns/features of the data which tend to be universal. This will be shown later in the results. 
\begin{figure}
	\centering
	\fbox{\includegraphics[width=0.8\linewidth]{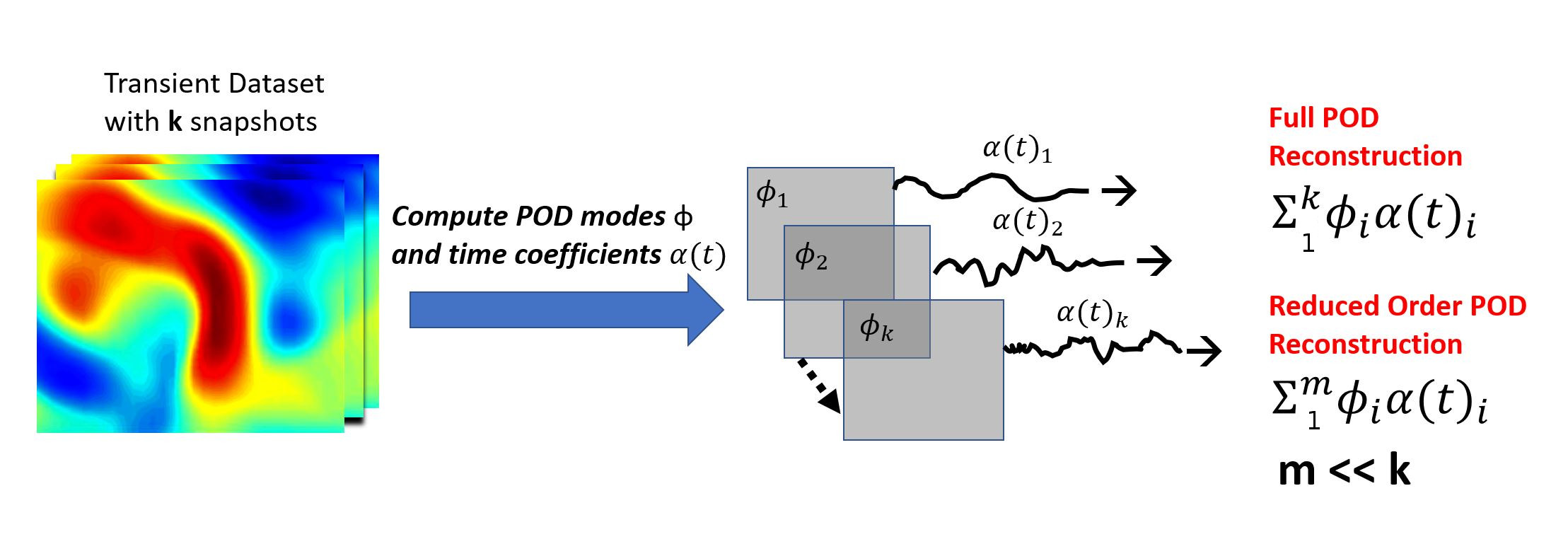}}
	\caption{Proper Orthogonal Decomposition and reduced order reconstruction}
	\label{POD}
\end{figure}
The essence of the LSTM-ROM approach is now outlined. A well-known strategy in literature to build ROMs involves using the dominant POD modes to represent
the key dynamics of the flow, since the dominant POD modes typically capture most of the flow energy~\cite{holmes2012turbulence}. Additionally, since the time
evolution of the POD modes is given by their temporal coefficients, they can be used to describe the temporal evolution of the key flow dynamics $f$, using Eqn~\ref{PODrecon}. A schematic is shown in Fig.~\ref{POD} for clarity. 

\begin{equation}
f \,=\, \sum_{i=1}^{i=k} \phi_{i}\alpha (t)_{i}
\label{PODrecon}
\end{equation}
where $\phi_{i}$ is a POD mode, with the flow consisting of $k$ total modes ($k \,=\,$ No. of snapshots). $\alpha (t)_{i}$ is the vector of time coefficients for a POD mode $\phi_{i}$. Subsequently, ROMs can be built by \textit{predicting $\alpha (t)_{i}$ at future time instants for dominant POD modes}, using a technique like GP. This approach is extremely cost effective for DNS/LES data, since it does not solve the full Navier-Stokes equations and instead models the time evolution of a few important flow structures. In this work, we retain this existing ROM framework, but explore LSTM NNs  to model $\alpha (t)_{i}$ instead of GP. The key steps in the LSTM-ROM methodology are outlined below and in a schematic in Figure~\ref{lstmrom}, while the details of NN training formulation will be explained in the subsequent paragraphs.

\begin{enumerate}
\item Select the number of 2-D planes to be used as training datasets (Figure~\ref{dataExtract})
\item From the planes, select a test dataset - this will be the dataset which will be modeled by the DL-ROM after it ``learns" the dynamics in the training datasets.
\item Extract dominant POD modes (usually 5-10 modes with the highest energy/eigenvalues) and their $\alpha (t)_{i}$ for each of the training datasets and test dataset. The $\alpha (t)_{i}$ of the test dataset POD modes will be used to validate the LSTM-ROM prediction.
\item Train LSTM/BiLSTM NN for the dominant POD modes chosen in step 3.
\item \textbf{Validation:} Using a short history of the test dataset POD mode $\alpha (t)_{i}$ as input, predict the next few time instants $\alpha (t + t^{\prime})_{i}$. Compare prediction with the true $\alpha (t + t^{\prime})_{i}$ from the test dataset. Repeat this for all the dominant POD modes chosen in step 3.
\item Using POD modes and the DL-ROM predicted temporal coefficients, compute predicted flow-field using Eqn.~\ref{PODrecon}.
\end{enumerate}

\begin{figure}
	\centering
	\fbox{\includegraphics[width=0.9\linewidth]{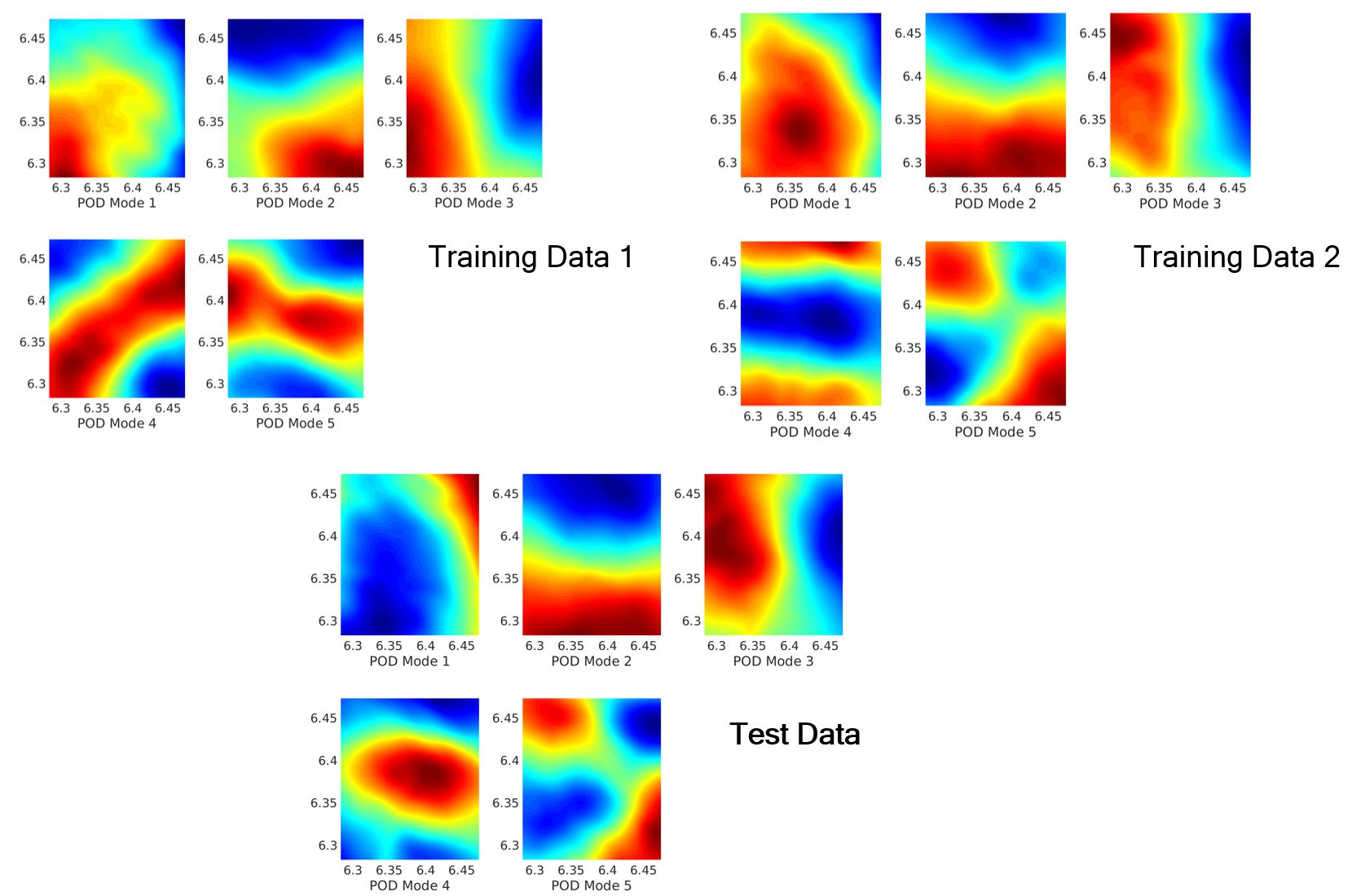}}
	\caption{Examples of Dominant POD mode U-velocity fields from training and test datasets}
	\label{podmodesIso}
\end{figure}

The details of the LSTM implementation software framework and training parameters can be found in the appendix. In this work, the POD modes and temporal coefficients from five 2-D planes which are equidistant from each other are used as the training data. The test dataset consists of a single 2-D plane also equidistant, whose POD temporal coefficients are modeled using the LSTM NN. An important assumption often made in ROM, including Galerkin-based ROM, is that the dominant POD modes for the training and test datasets are qualitatively similar~\cite{noack2011reduced}. For instance, flows within a narrow range of Reynolds number can exhibit qualitatively (but not quantitative) similar behavior, which are encoded in their dominant POD modes~\cite{noack2011reduced}. For the ISO training and test 2-D planes used here, the dominant 5 POD modes for some of the training datasets and test data are shown in Fig.~\ref{podmodesIso}, for reference. The modes show significant qualitative similarity, although differences exist. This is indeed amenable to data-driven techniques like deep learning, since 
the goal is to extract key patterns which are common in the dataset, and use them to model behavior of an \textit{unseen} dataset of the same family. 
\begin{figure}
	\centering
	\fbox{\includegraphics[width=0.9\linewidth]{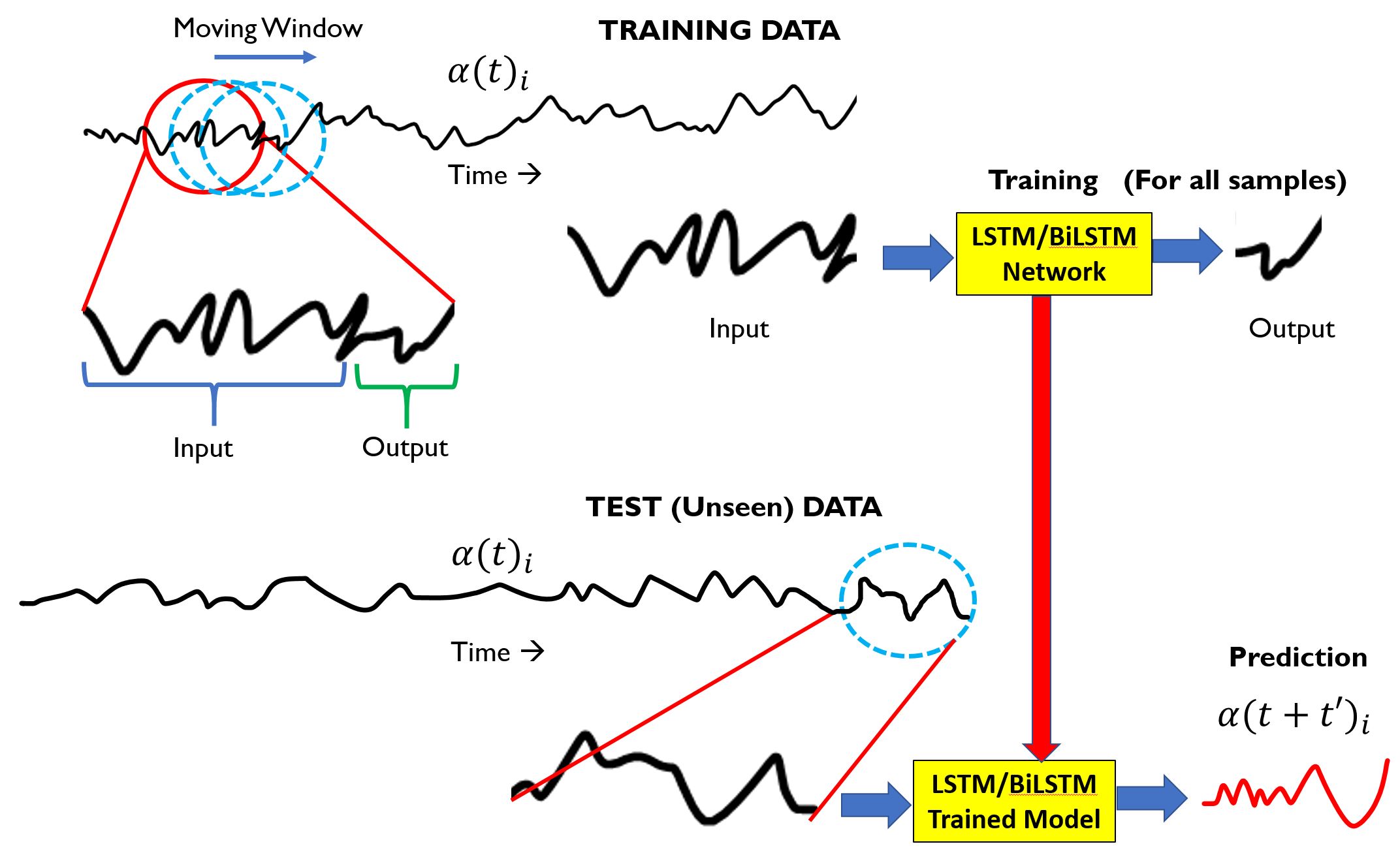}}
	\caption{LSTM training strategy with input/output framework}
	\label{lstmprediction}
\end{figure}
Since the desired goal of the LSTM-ROM is to predict $\alpha (t + t^{\prime})_{i}$ given a $\alpha (t)_{i}$, we propose that the training strategy can be framed as an input/output problem.
The training strategy is explained in the Fig.~\ref{lstmprediction} schematic.
The training $\alpha (t)_{i}$ signals from all the datasets are decomposed into several short samples by a moving window. These samples are divided into two parts - an input and its corresponding output. The LSTM NN is trained by providing it a set of input and output signals. For example, a training sample will contain an input signal of length 0-0.1 seconds, with the output signal being 0.1-0.2 seconds. Therefore, the time window is 0.1 seconds and the prediction horizon is also 0.1 seconds, with the LSTM NN learning the associative relationship between thousands of such input-output pairs from all the training datasets. In essence, the training problem now boils down to learning signatures in 1-D signals from a \textit{training dataset}, and using the learned patterns to predict realizations of 1-D signal from a test dataset, given the previous time instants as an input to the NN. After training is satisfactorily complete, a prediction is made by giving an unseen input of 0.1 seconds from the test dataset and the model outputs a 0.1 second signal. 

\section{Results}
The LSTM architecture described above is trained with different lengths of the prediction horizons and time windows. The time window is an indicator of the ``memory" in the flow and while its exact computation is not feasible, we will explore an approximate approach in sections ahead. These time window/prediction horizon choices have to be typically made at the NN design level before the training process begins, as they can significantly affect the accuracy of the model. Therefore, some trial and error is involved in making these choices, based on the physics of the flow. The results presented in this section have been generated with a time window of 10 time steps and a predictive horizon of 10 time steps. This single, modest value for the horizon is chosen so as to make the NN training computationally manageable, and to demonstrate the efficacy of this method. We investigate the two turbulence datasets above to illustrate different training paradigms using the LSTM architectures described in the previous section. We will study the performance with using longer horizons (and equally long time windows) in the forthcoming section.

\subsection{Isotropic Turbulence}
\begin{figure}
	\centering
	\fbox{\includegraphics[width=0.8\linewidth]{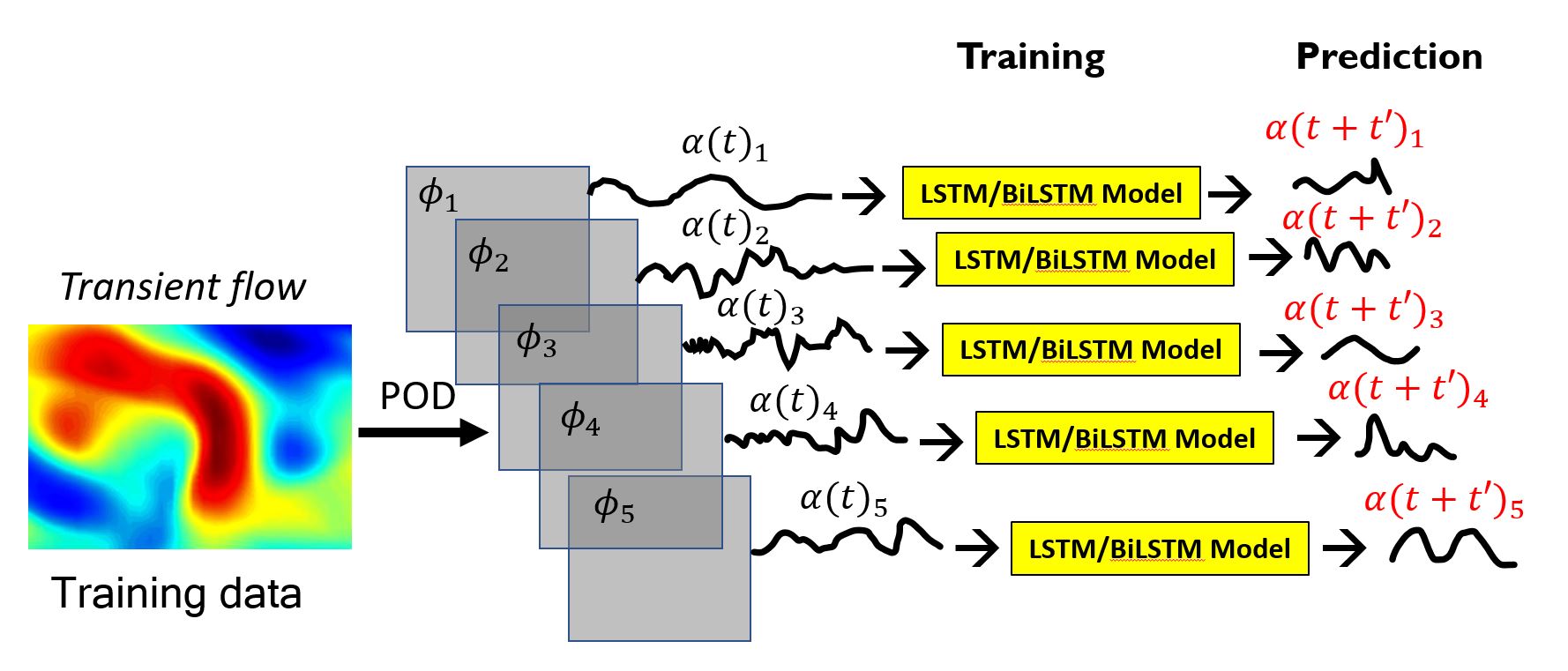}}
	\caption{Training of a Multiple NN models for all POD dominant modes}
	\label{multiplemodel}
\end{figure}

\begin{figure}
	\centering
	\fbox{\includegraphics[width=0.75\linewidth]{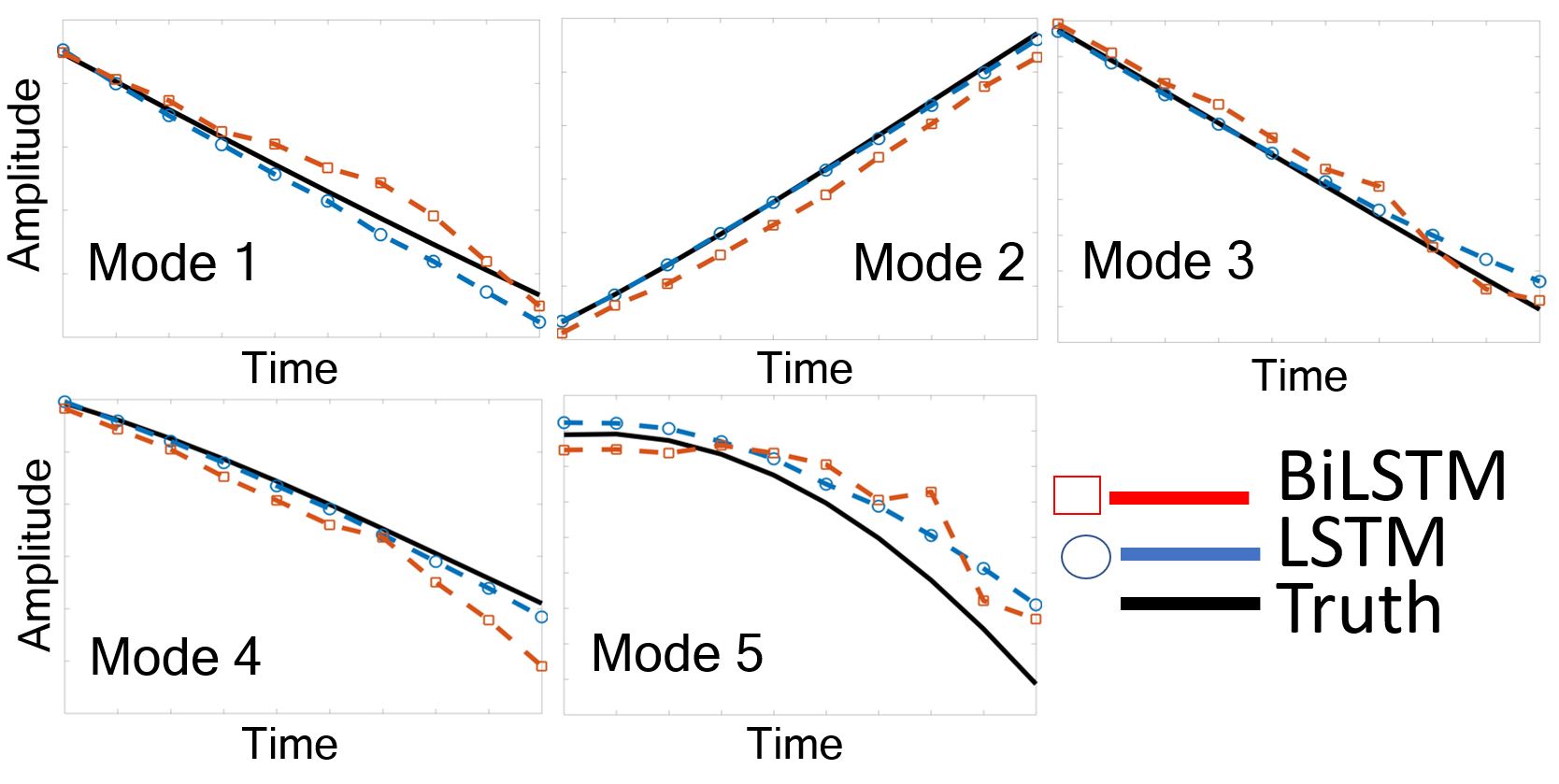}}
	\caption{LSTM and BiLSTM predictions of Dominant POD $\alpha (t+t^{\prime})$ for Isotropic turbulence test data}
	\label{isoresults}
\end{figure}

The isotropic turbulence dataset is decomposed into 2D slices (planes) as mentioned in the previous section and its temporal coefficients are used to train the NN. A direct approach to performing this is to train a) LSTM, and b) BiLSTM on each of the $5$ dominant POD temporal coefficients, such that we obtain $5$ trained models - one for every mode (as shown in the schematic in Fig.~\ref{multiplemodel}). For instance, there is a dedicated model for POD mode 1, which is then used to predict the POD mode 1 temporal mode coefficients for the test dataset. Therefore, if we build a ROM using dominant $m$ modes, we would need $m$ models. 
In order to evaluate the prediction, we provide an input to the model \textit{any sample} from the temporal coefficients of the relevant mode from the test dataset. As mentioned before, the entire temporal coefficient 1-D signal has been divided into thousands of samples of equal length. Upon receiving any of these samples as input, the NN model should predict the sample which follows immediately after it. This test, although simple; can shed considerable light on how capable the LSTM based NNs are, since we have thousands of samples to test for a given mode, making the findings statistically significant. The prediction results at a randomly chosen sample for all $5$ dominant POD modes are provided in Fig.~\ref{isoresults}.
The figures show 3 curves -  the expected (truth) data, the LSTM NN prediction and the BiLSTM NN prediction.  It is evident that the predictions are following expected trends, especially with LSTM NN. Interestingly, the BiLSTM architecture appears to be less accurate than the LSTM, which is intriguing considering the theoretical justification behind its architecture and success in literature. Possible reasons for this discrepancy will be explored shortly. In order to study the accuracy of these NN architectures it is necessary to gauge the prediction error over a large number of samples, for statistical significance of our observations. The Mean Absolute Scaled Error~\cite{hyndman2006another} (MASE) metric is chosen to quantify the deviation of prediction from the expected trend. 

\begin{figure}
	\centering
	\fbox{\includegraphics[width=1\linewidth]{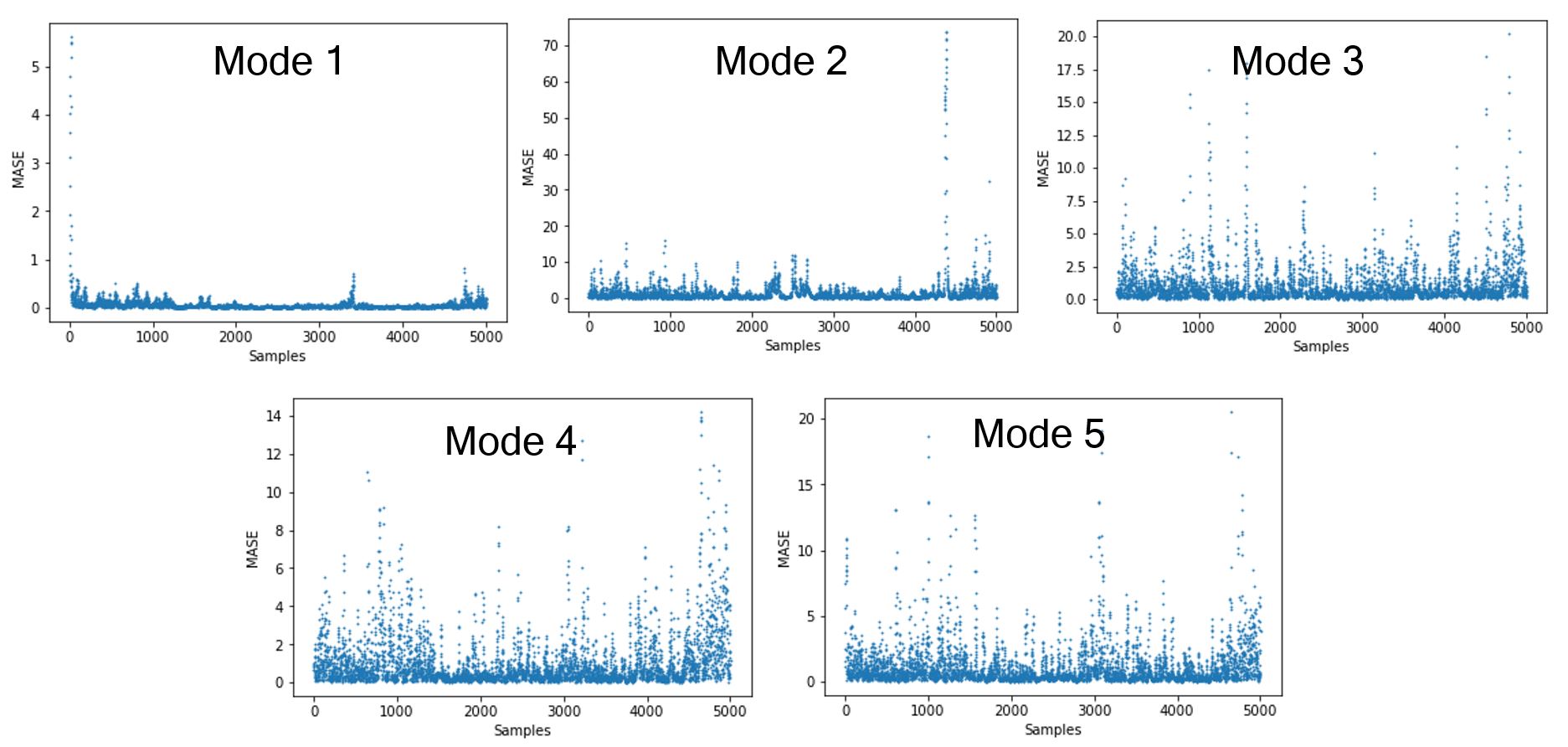}}
	\caption{Mean Absolute Scaled Error (MASE) for LSTM predictions on all test samples in ISO dataset}
	\label{MAEisoLSTM}
\end{figure}

\begin{figure}
	\centering
	\fbox{\includegraphics[width=1\linewidth]{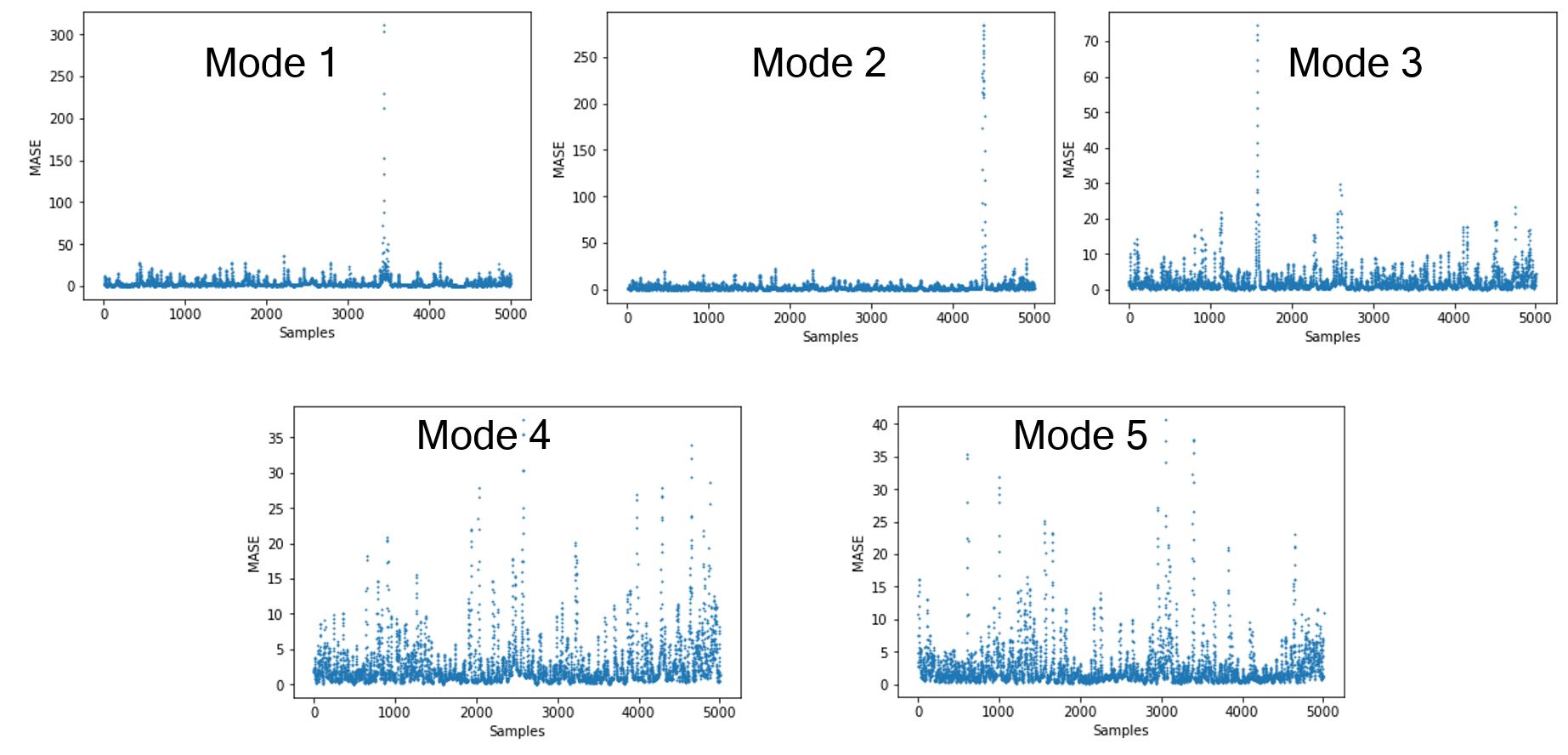}}
	\caption{Mean Absolute Scaled Error (MASE) for BiLSTM predictions on all test samples in ISO dataset}
	\label{MAEisoBiLSTM}
\end{figure}

\begin{figure}
	\centering
	\fbox{\includegraphics[width=0.6\linewidth]{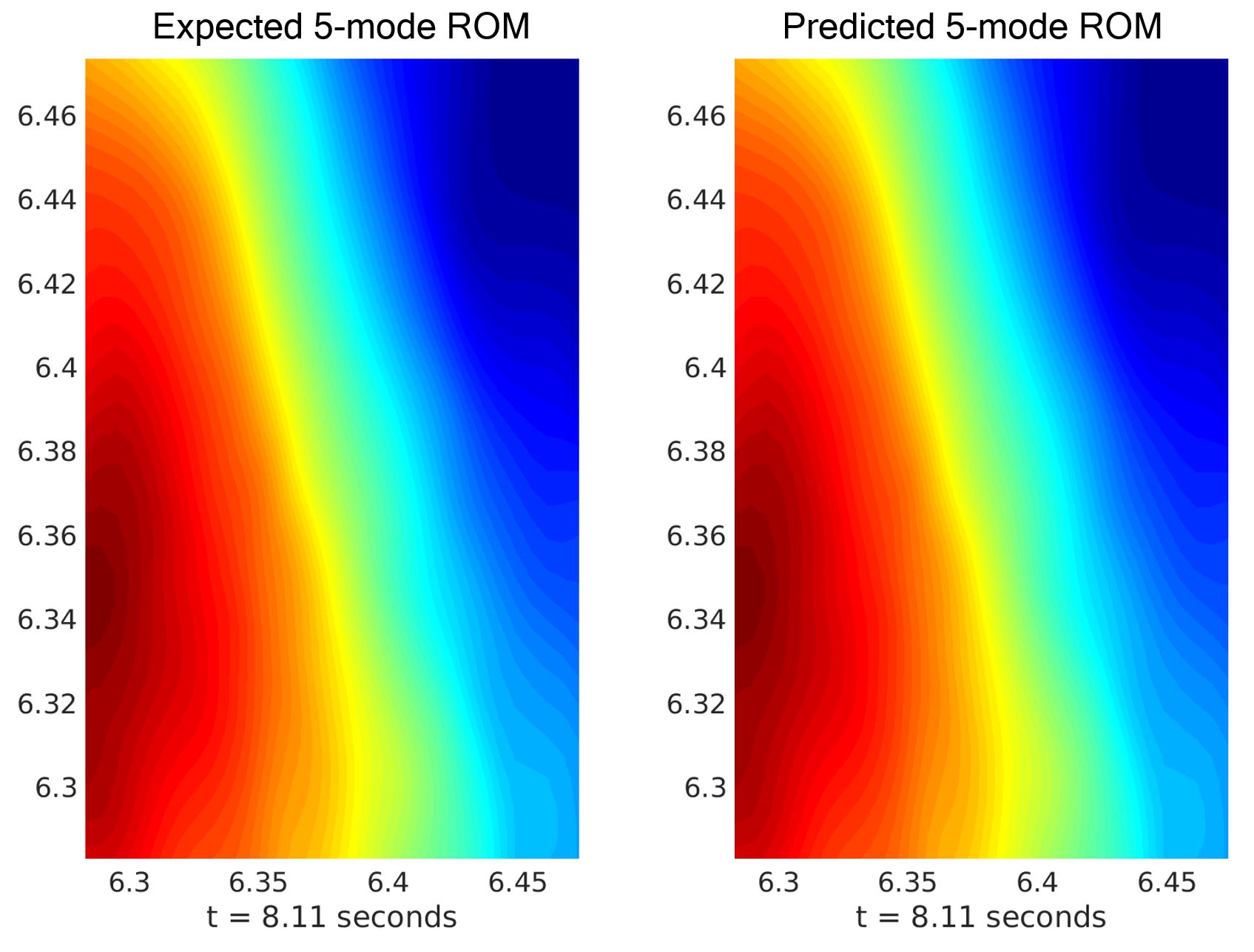}}
	\caption{Predicted 5-mode LSTM-ROM flow with actual 5-mode POD reconstructed flow at $t\,=\,8.10$}
	\label{pred1}
\end{figure}

Figure~\ref{MAEisoLSTM} shows the MASE of the LSTM predictions for all samples in each of the $5$ dominant POD modes, with the average MASE for that mode. With a prediction horizon of $10$, a total of $5003$ samples can be generated for each POD temporal coefficient, since it has a total of $5023$ realizations. The results show that the MASE is generally low, except at samples where a sudden increase is observed. A similar trend is also observed for BiLSTM in Fig.~\ref{MAEisoBiLSTM}, although the average MASE is higher than LSTM across all $5$ POD modes. Finally, the predicted temporal coefficients can be used to compute future time evolution of the flow using Eqn.~\ref{PODrecon}. A comparison of the predicted flow field with the expected flow at a given time instant is shown in Fig.~\ref{pred1}. Since dominant modes comprise a significant amount of flow energy, prediction errors in the lower modes (like POD mode 4 and 5) tend to less negatively impact the flow field accuracy  than errors in POD modes 1 and 2. Naturally, a question arises about the distribution of errors with various POD modes, and this will be discussed in the next section. 
Now we discuss the lower prediction accuracy in BiLSTM, which initially appears counter-intuitive. Our explanation for this behavior stems from the fact that BiLSTM was explicitly developed for long range dependencies in the signals, where a forward and backward pass would help the NN learn the input-output relationships (Fig.~\ref{lstmcomparison}) better. As mentioned previously, the key application area of BiLSTM appears to be language modeling for complex tasks, where well-defined patterns have been observed in literature. Some of the most common examples are presence of high-frequency words in the English language texts, mentioned in the seminal work on redundancies and statistics by Claude Shannon~\cite{shannon1951prediction}. Additionally, literature in information theory and computational linguistics point to significant long range statistical correlations in multiple languages~\cite{hatzigeorgiu2001word,li1992random,calude2011we}. As such, it is fair to assume that most published English language text follows standard grammar, which makes it likely for two words in a sentence, separated by several other words, to have a strong correlation. We surmise that this could be a reason why the forward and backward pass training strategy in BiLSTM has been shown to improve performance considerably when compared to LSTM, where only a single forward pass is made.
\begin{figure}
	\centering
	\fbox{\includegraphics[width=0.4\linewidth]{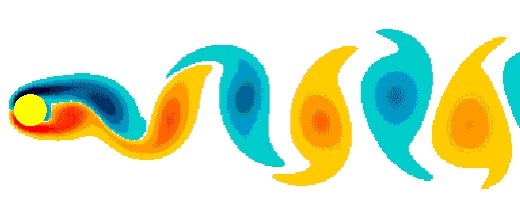}}
	\caption{Vortex Shedding behind a cylinder, displaying strong correlations in spatio-temporal wake dynamics}
	\label{shedding}
\end{figure}
In contrast, for signals generated from highly chaotic, nonlinear dynamical systems like turbulence, presence of long range correlations is not guaranteed in all applications, though there might be exceptions. In such scenarios, it is likely that BiLSTM overfits the data and learns dependencies that do not exist, leading to lower predictive accuracy seen in the results here. It should be noted that situations do exist where long range correlated behavior has been observed in turbulent flows, such as bluff body wake shedding (Example shown in Fig.~\ref{shedding}). In this flow, a well placed probe in the trailing wake region may capture periodic and quasi-intermittent features over time. In such cases, the BiLSTM algorithm may be a good candidate. While more studies would certainly be necessary to better analyze BiLSTM performance for a range of chaotic systems, they are outside the scope of this work.

\subsection{Magnetohydrodynamic Turbulence}

The strategy in the previous section required a NN model to be trained for each POD mode - a \textit{multiple model approach}. This approach implies that the NN learns universal features \textit{for the same mode} between the various training datasets. The implicit assumption in this approach is that the features to be learned in, say Mode 1, are only present in the corresponding Mode 1 in the training datasets. However, modal decompositions like POD are a linear combination of eigenvectors and eigenvalues, which do not explicitly account for the inter-modal, nonlinear interactions seen in turbulence. A classic example of this is the energy transfer from larger to smaller scales (the energy cascade) and the less significant mechanism of the inverse energy cascade (or backscatter) from smaller to larger scales. Therefore, instead of the flow dynamics being artificially constrained to a single POD mode, in reality the signature of a flow feature (like a vortex) may be spread across a range of POD modes. As a result, the multiple model approach may not capture the full signature of the flow dynamics in a POD mode, since it learns from only the training dataset POD modes of the same rank.
Another practical issue with the multiple model approach is the increased memory requirements for embedding NN models in the on-board electronic hardware for flow control. Increasing the fidelity of the ROM by using more POD modes can cause a memory bottleneck since $N$ modes would require $N$ models. 

\begin{figure}
	\centering
	\fbox{\includegraphics[width=1\linewidth]{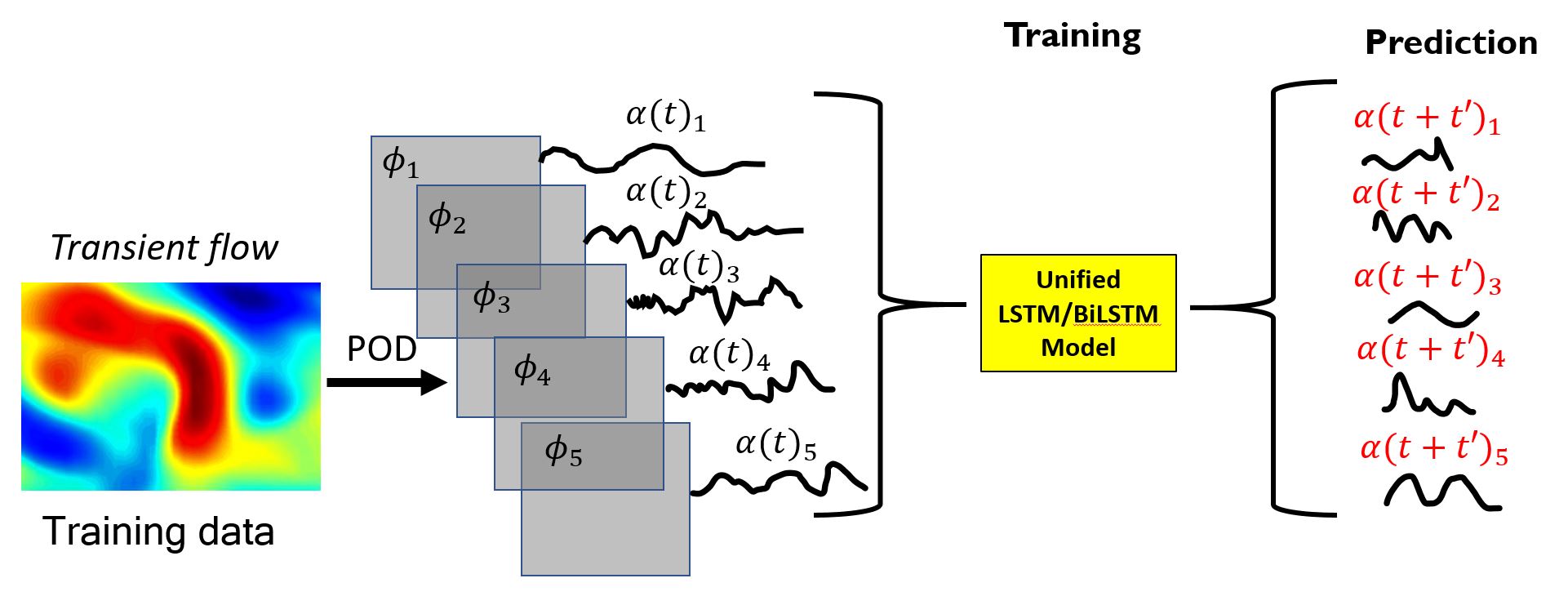}}
	\caption{Training of a unified NN model for all POD dominant modes}
	\label{unifiedmodel}
\end{figure}

In this section, we propose an alternative training strategy to account for these issues - the \textit{unified model approach}. In this approach, the NN is trained with samples not just from POD modes of the same rank, but including other ranks as well. For instance, the LSTM (or BiLSTM) NN will be trained with all samples from POD mode 1,2,3,4 and 5, from all training datasets. This results in a single, \textit{unified} NN model, which can be used to predict the coefficients for all the POD modes. The schematic in Fig.~\ref{unifiedmodel} outlines this approach.
\begin{figure}
	\centering
	\fbox{\includegraphics[width=0.75\linewidth]{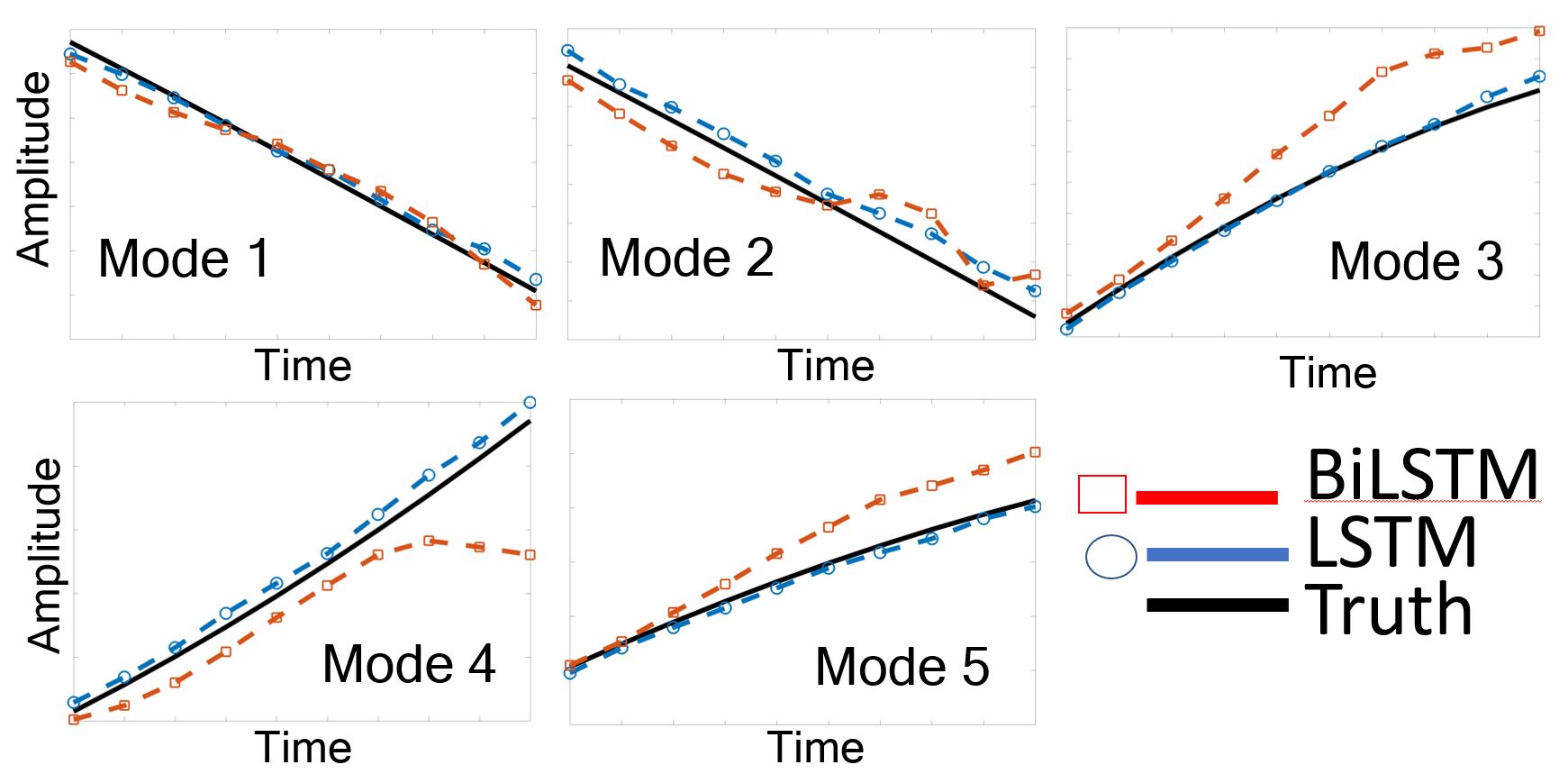}}
	\caption{LSTM and BiLSTM predictions of Dominant POD $\alpha (t+t^{\prime})$ for MHD test data}
	\label{mhdresults}
\end{figure}

\begin{figure}
	\centering
	\fbox{\includegraphics[width=1\linewidth]{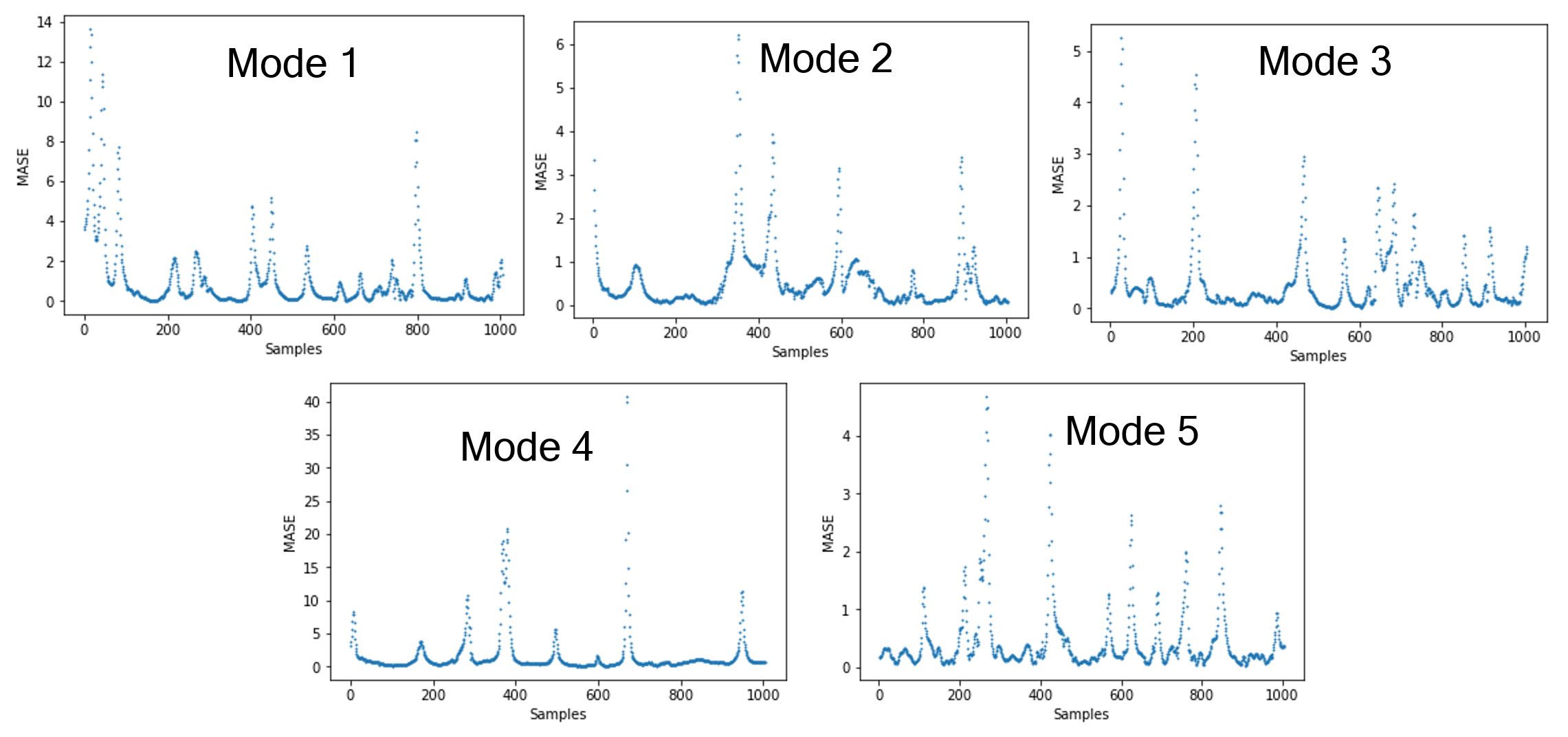}}
	\caption{Mean Absolute Scaled Error (MASE) for LSTM predictions on all test samples in MHD dataset with Unified Model}
	\label{MHDMAELSTM}
\end{figure}

\begin{figure}
	\centering
	\fbox{\includegraphics[width=1\linewidth]{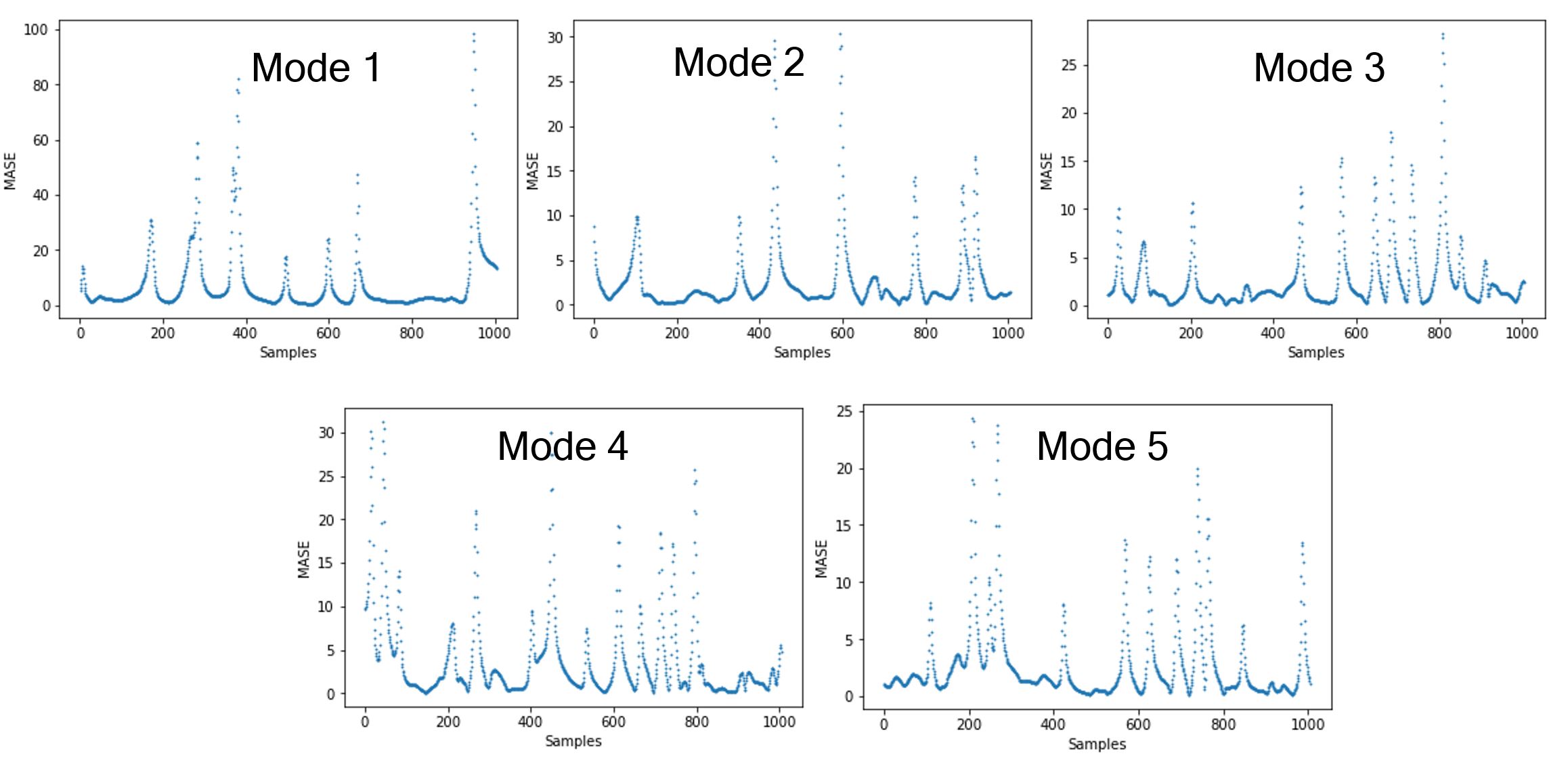}}
	\caption{Mean Absolute Scaled Error (MASE) for BiLSTM predictions on all test samples in MHD dataset with Unified Model}
	\label{MHDMAEBiLSTM}
\end{figure}
The LSTM and BiLSTM NNs are trained just like the previous section, but using the unified model approach. The results at a randomly chosen sample are shown Fig.~\ref{mhdresults}, indicating that the LSTM model is able to predict the data with considerable accuracy for each of the modes individually. We again plot the MASE for LSTM and BiLSTM to quantify the NN performance for all samples, which is shown in Fig.~\ref{MHDMAELSTM} and Fig.~\ref{MHDMAEBiLSTM} respectively. Again, we generally see good performance from the unified model for most samples. Therefore, \textit{it is possible that the unified model is learning statistics common among different POD modes}, thus boosting performance without the need for building separate models. Additionally, we once again see that the BiLSTM consistently under-performs the LSTM predictions - further strengthening our hypothesis in the previous section. 

\section{Memory effects and LSTM Model Accuracy}

The primary focus of LSTM (and its variants) is to model sequential datasets by extracting the relationship i.e. correlations between subsequent realizations. In essence, LSTM \textbf{assumes that there is memory in the sequence}. However, signals originating from chaotic dynamical systems are known to have very short correlated events and memory does not typically persist over long time periods. Yet, in some cases (like Fig.~\ref{shedding}) persistent memory might be a reasonable assumption. 

\begin{figure}
	\centering
	\fbox{\includegraphics[width=0.75\linewidth]{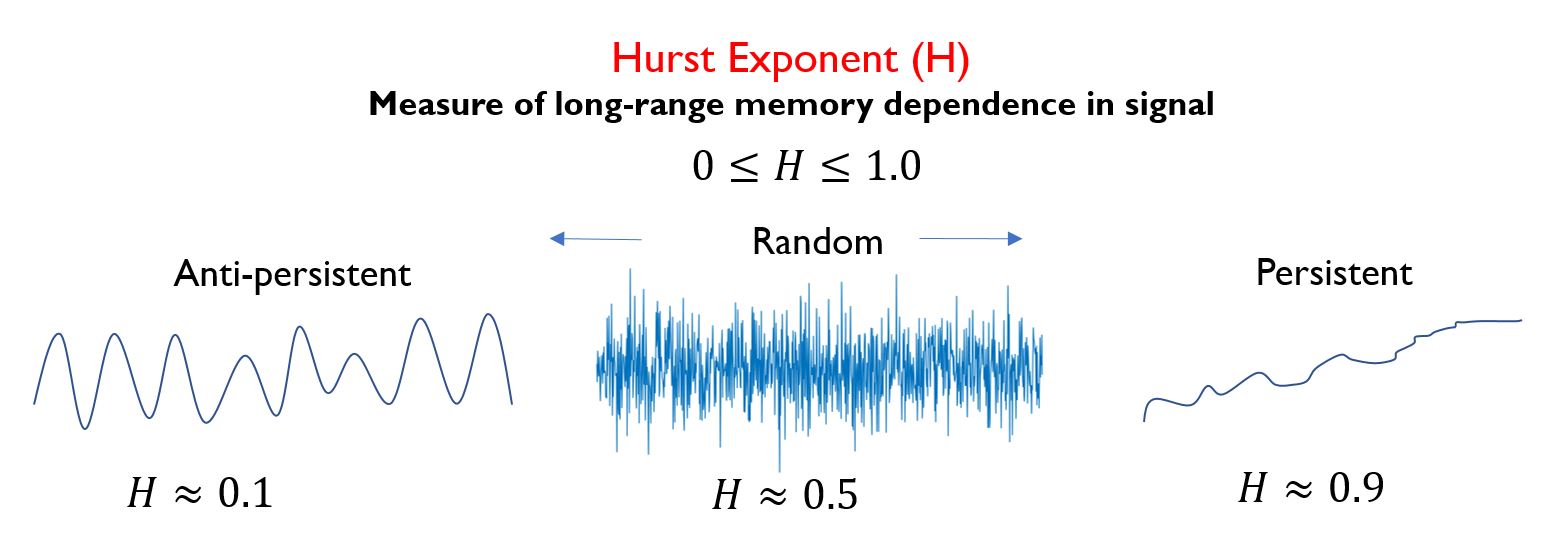}}
	\caption{The Hurst Exponent}
	\label{hurstSchematic}
\end{figure}

A question now arises: \textit{Can we quantify ``memory" in a sequential dataset, such that the suitability of LSTM as a predictive modeling approach can be evaluated?} 
In this section, we explore the relationship between the theoretical underpinnings of LSTM and the chaotic nature of the training data, to study their effectiveness in modeling turbulence, as seen in Figs~\ref{isoresults} and~\ref{mhdresults}.
While the concept of memory in a sequence can be intuitive, quantifying it is not straightforward. A prominent solution in literature is the \textbf{Hurst Exponent}, which was developed by E. Hurst while studying water level fluctuations of the Nile river~\cite{hurst1951long} and subsequently popularized by Mandelbrot~\cite{mandelbrot1968noah}. The Hurst exponent is agnostic of the dataset, and have been used extensively in hydrology~\cite{koutsoyiannis2003climate}, finance~\cite{carbone2004time}, climate sciences~\cite{peng2012trend} and genetics~\cite{roche2003long}.
The Hurst Exponent, $H$, is a quantitative estimate of the presence or absence of long-term trends in a sequential one-dimensional signal, like a time series. It is derived from rescaled-range analysis, which itself is a measure of how variable a time series for different lengths, using the ratio of its range and the standard deviation. Hurst exponent can be statistically expressed as,

\begin{equation}
    \mathbb{E}\left[\frac{R(k)}{S(k)}\right] \,=\, C k^{H}  
\end{equation}
as $k \longrightarrow \infty $.

Here, $R(k)$ is the range of the first $k$ values in the series, and $S(k)$ is the corresponding standard deviation. $\mathbb{E}$ is the expected value of the ratio, with $k$ being the number of data points in the series being currently processed. $C$ is a constant, and the exponent $H$ is the Hurst exponent. $H$ lies between 0 and 1, as shown in the schematic in Fig.~\ref{hurstSchematic}. A $H \to 0$ indicates \textbf{anti-persistent} behavior i.e. an upward trend in the sequence is most likely to be followed by a downward trend (and vice versa), while $H \to 1$ indicates a \textbf{persistent} trend i.e. an increase is most likely followed by another increase. However, the most useful component of this metric is when $H \,=\, 0.5$, indicating purely \textbf{random} behavior - in this regime, the trend behaves as a random walk (white noise, brownian motion etc.). $H$ values for any sequence closer to $0.5$ indicate the more randomness, and consequently less discernible trends. \textit{Therefore, $H \approx 0.5$ implies the lack of clear persistent/anti-persistent trends in the sequence i.e. lack of memory}. This has important consequences with regards to predictive modeling using LSTM, which has the assumption of memory intrinsically built into the algorithm.  
\begin{figure}
	\centering
	\fbox{\includegraphics[width=0.8\linewidth]{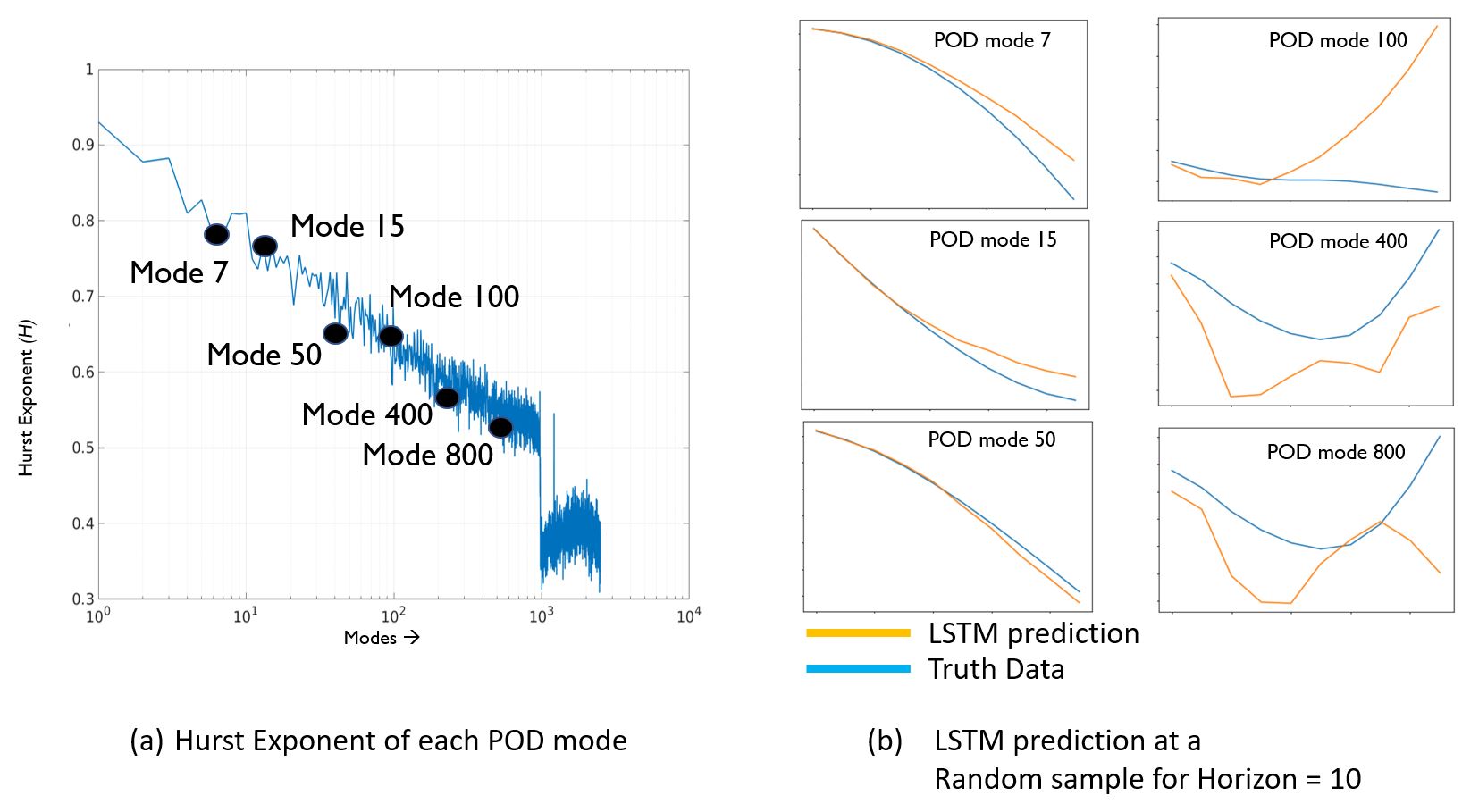}}
	\caption{Memory in various POD coefficients as estimated by the Hurst Exponent for the ISO dataset}
	\label{hurstResultsISO}
\end{figure}
\begin{figure}
	\centering
	\fbox{\includegraphics[width=0.8\linewidth]{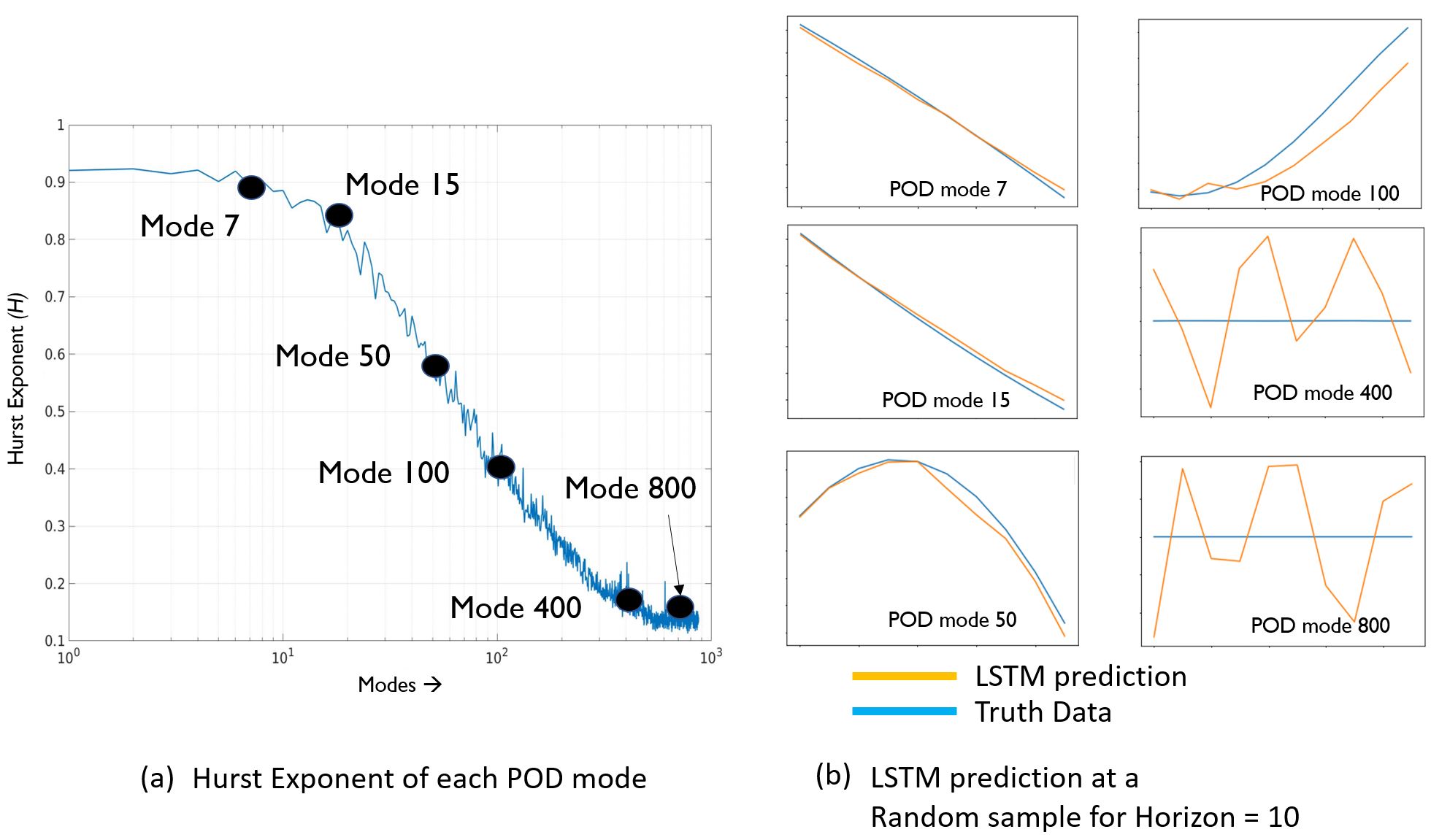}}
	\caption{Memory in various POD coefficients as estimated by the Hurst Exponent for the MHD dataset}
	\label{hurstResultsMHD}
\end{figure}
One of the major factors influencing the accuracy of the LSTM-ROM is the interplay between \textbf{persistence} and \textbf{horizon}. Persistence refers to the behavior of a sequential series described by the Hurst Exponent, and is indicative of the ``memory'' in the signal. Horizon is the number of steps ahead that we want a model (in this case, LSTM) to predict. In this section, we will study how persistence influences the accuracy of the prediction horizon. In order to do so, we again utilize the ISO and MHD datasets with the LSTM-ROM methodology. We first estimate $H$ for the $\alpha(t)$ of all the POD modes (with non-negligible eigenvalues) from the dataset, following which we choose modes in different $H$ regimes i.e. \textbf{persistent}, \textbf{random} and \textbf{anti-persistent}. We then develop LSTM models for each of these modes, for a given horizon length. In this regard, the analysis in this section differs from the previous sections in two major ways: a) Previously, the LSTM training hyper-parameters were tuned for only one (the dominant) mode and then used to train all other modes, to save computational cost. In this section, we tune the LSTM NN for each mode. This ensures that we obtain the best model for any given mode, which is necessary to establish the best case performance using LSTMs for that mode. b) While the horizon length was constant at $L \,=\,10$, we now vary $L$ and repeat the LSTM-ROM approach for each $L$. Doing so will give us valuable insight into its relationship with accuracy and persistence.
\begin{figure}
	\centering
	\includegraphics[width=1\linewidth]{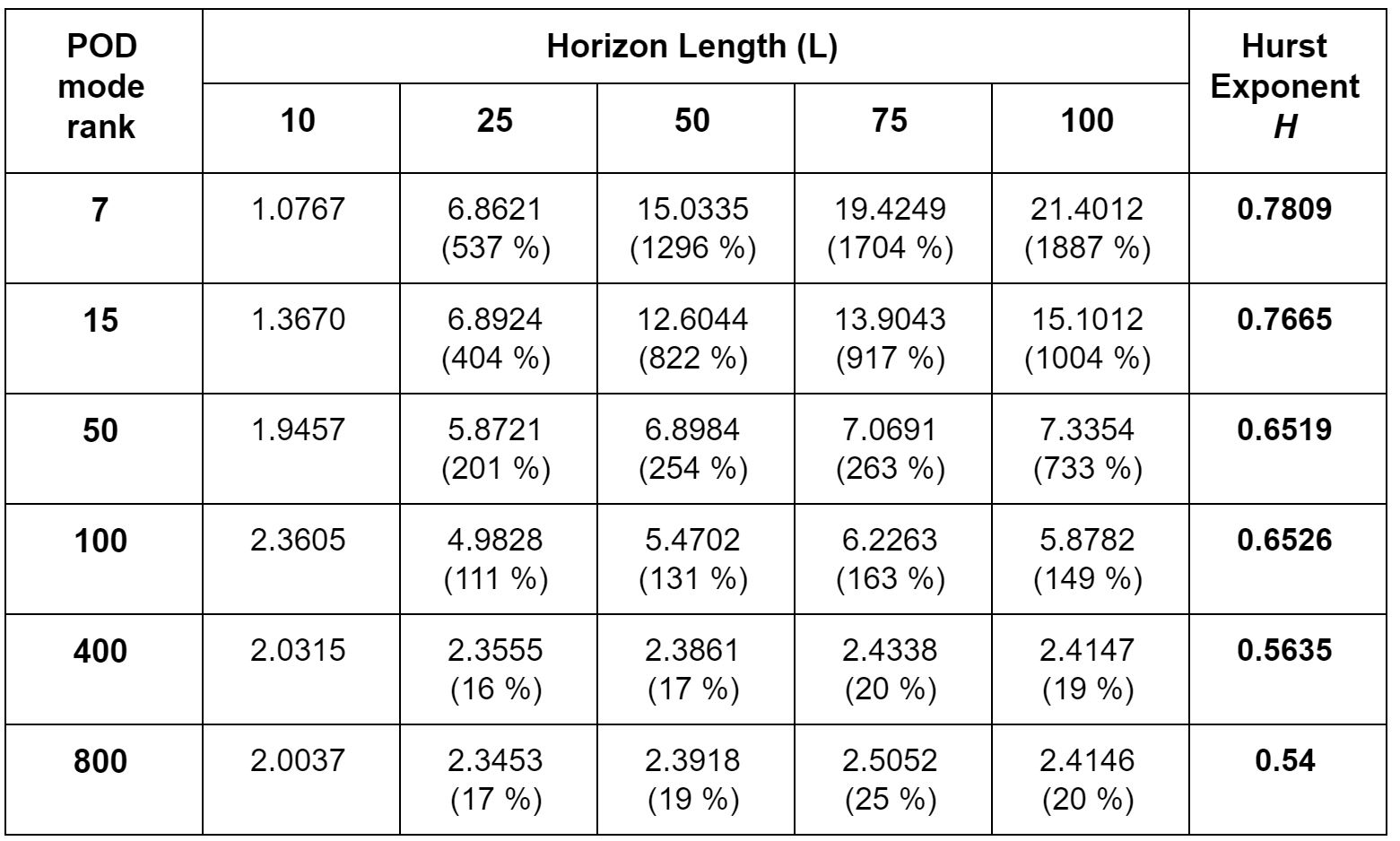}
	\caption{Impact of Persistence and Horizon on LSTM-ROM model accuracy in ISO dataset}
	\label{ISOtable}
\end{figure}
\begin{figure}
	\centering
	\includegraphics[width=1\linewidth]{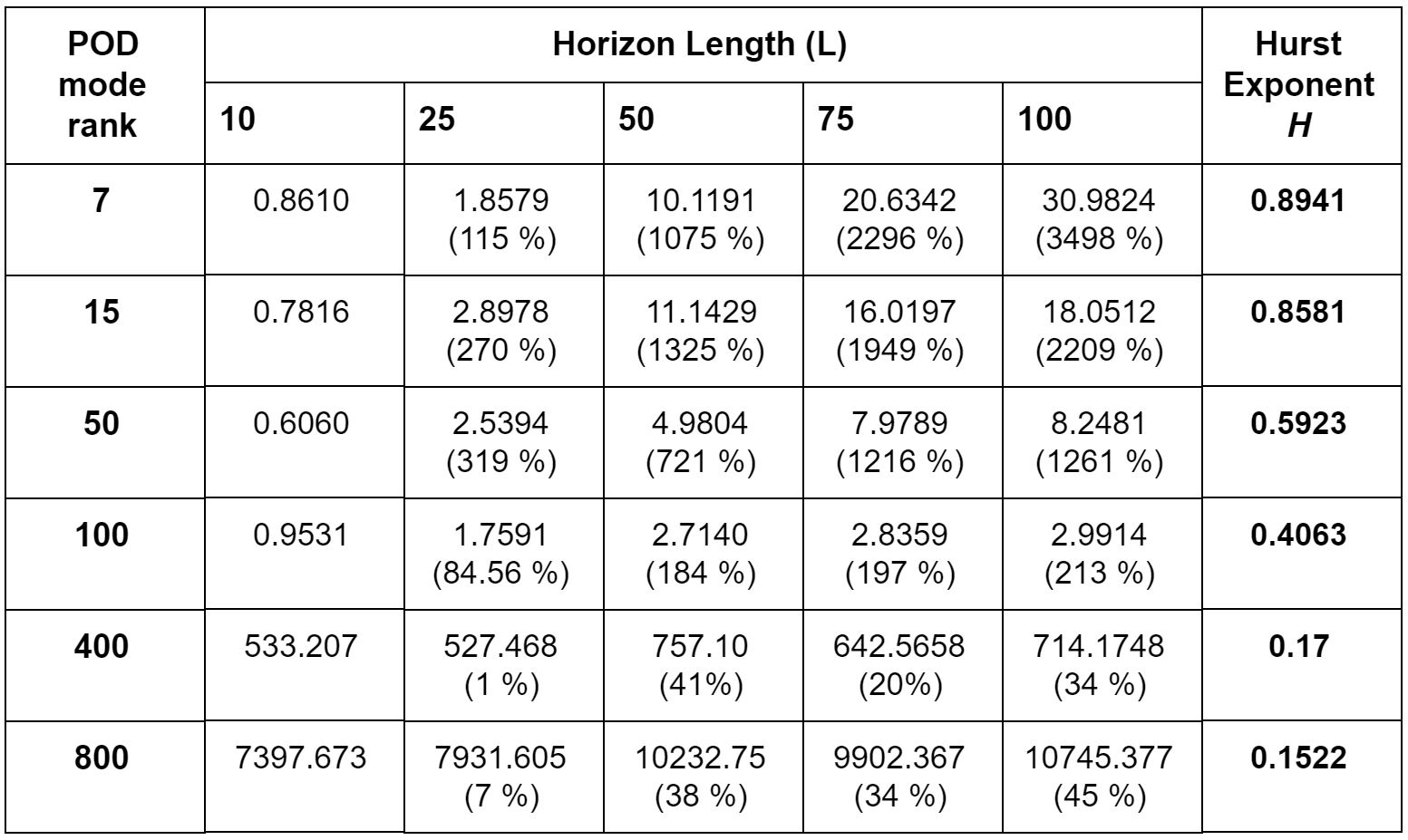}
	\caption{Impact of Persistence and Horizon on LSTM-ROM model accuracy in MHD dataset}
	\label{MHDtable}
\end{figure}
Consider now the data for ISO and MHD datasets in Tables~\ref{ISOtable} and \ref{MHDtable} respectively. The tables display the persistence and horizons for some POD modes, along with the average accuracy of the predicted models, in terms of the mean MASE. Additionally, a helpful representation is the \textbf{percentage change in mean MASE}, which is a measure of how the error increases/decreases compared to $L \,=\, 10$ as a baseline. For instance, in POD mode 7 we see that percent increase in error for $L \,=\, 25$ jumps to $537 \%$ and further increases by at least an order of magnitude for higher $L$ i.e. $50$, $75$ and $100$. This indicates that the LSTM-ROM performance deteriorates rapidly with an increase in the prediction horizon - a pertinent observation since the LSTM hyper-parameters were optimized individually at each $L$ for maximum accuracy. While this is true for a single mode, a clear picture emerges when this analysis is repeated for \textit{all} $L$.

Consider the data in Tables~\ref{ISOtable} and \ref{MHDtable}. While the mean MASE clearly increases with horizon for all modes, the magnitude of increase in error is the highest for the POD modes $7$ and $15$, as shown by the percentages.  A less drastic variation is observed for POD modes $50$ and $100$, especially for $L = 50-100$. Finally, low rank modes  $400$ and $800$ show very different trends, with the mean MASE undergoing very minimal variation from the baseline at $L = 10$, meaning that the prediction error stays consistently high regardless of the horizon length. Interestingly in Table~\ref{MHDtable}, these modes show highly \textit{anti-persistent} behavior and modeling accuracy is extremely poor for all horizons, especially considering how accurate LSTM is in modeling highly \textit{persistent} behaviors. At this outset, a few observations can be made about the LSTM behavior:

\begin{enumerate}
	\item Strongly persistent behavior (as shown by low rank modes) tends to be modeled accurately for short horizons, with drastic decrease in accuracy for longer horizons.
	\item Conversely, LSTM has considerable difficulties modeling strongly anti-persistent behavior, regardless of horizon.
	\item For any given mode, increase in horizon tends to reduce accuracy, with the effect being less pronounced for strongly anti-persistent modes.
	\item For weakly persistent modes i.e. those with $H \to 0.5$, there may be some improvement in accuracy with increase in horizon.
\end{enumerate}

We can see that the highly persistent modes tend to be low rank, containing more energy than high rank modes. Likewise, weakly-persistent and highly anti-persistent modes tend to have low energy. As a result, the behaviors observed above tend to have a direct effect on the quality of a turbulence ROM. For short horizons, accurate modeling of the high energy modes leads to accurate ROMs, while their poor modeling accuracy at longer horizons leads to inferior ROMs. This is true, even though low energy modes have marginally better accuracy at long horizons, as seen in the tables above.

\section{Discussion \& Conclusions}
This paper proposes a deep learning based ROM  for turbulent flows for flow control applications using the Long Short Term Memory (LSTM) neural network , since they have demonstrated immense potential in modeling complex sequential data in other domains. We now outline some merits, limitations for the LSTM-ROM and avenues for further improvement based on our analysis so far.
One of the more interesting observations in this work was that Bidirectional LSTM consistently performed worse than the traditional LSTM, despite its theoretical formulation intending otherwise. We surmise that this is likely because it over-fits data, by assuming long range memory that may not have actually existed. While LSTM was more accurate, it was seen that its accuracy deteriorated with an increase in horizon. While we made every effort in this work to tune the neural network hyper-parameters to improve accuracy, there is a possibility that a further improvement could have been obtained. However, we believe that any such gains would have been marginal and the qualitative trends would hold. Furthermore, our accuracy may also be theoretically restricted due to the \textbf{Lyapunov exponent theory} for dynamical systems~\cite{sprott2003chaos}. The Lyapunov exponent $\lambda$  is defined in Eqn.~\ref{lyapunov} as,

\begin{equation}
|\delta F(t)| \approx e^{\lambda t} |\delta F(t)_{0}|
\label{lyapunov}
\end{equation}

Where $|\delta F(t)|$ signifies the separation in trajectories of a dynamical system. The Lyapunov exponent shows that the farther in time we move from the initial state $|\delta F(t)_{0}|$, the greater the rate of divergence between trajectories that were closer at $t \to 0$. Intuitively, in a dynamical system the likelihood of making accurate predictions for time series farther away from the origin, drops exponentially. However, \textit{this merely indicates the accurate,  pure data-driven approaches to prediction of long time horizons may have several difficulties}. Developing  governing equation, physics-based models (Navier Stokes equations) with a data-driven approach (LSTM) may still provide us with an accurate and efficient prediction scheme that complements the strengths of both approaches, while not being constrained by the Lyapunov exponent. 

Additionally, we must address the implicit assumption made here, that the dominant POD spatial modes are consistent within the same regime. This assumption has been used in literature extensively, especially when using Galerkin projection based ROM equations for Navier Stokes. This generally holds for simplified flow fields and geometries, like cylinder wake shedding at very closely related Reynolds numbers (as long as they do not undergo bifurcations). However, this can be extremely restrictive for problems with large variations in Reynolds numbers, since such flows exhibit both spatial and temporal dynamics. In this regard, future work would focus on exploring LSTM based modeling of turbulence physics, augmented with the Navier-Stokes equations to enforce the spatial dynamics and symmetries associated with turbulence.

An overarching theme of our effort is to demonstrate that capability of LSTMs in modeling non-stationary signals from high-fidelity turbulence; thus the pedagogical nature of this work. A suitable application of the data-driven LSTM-ROM approach described in this work would be in specialized engineering flow control applications where the model has to generalize to a narrow class of regimes and a ``physics" insight into the flow is not a necessity. Additionally, an added advantage of these trained LSTM-ROM models is that they have a very low computational cost for inference, which is convenient for resource-scarce on-board control hardware like micro-controllers and devices with limited memory. An additional contribution of this work is utilizing Hurst exponent is a useful tool to study the applicability to predictive modeling approaches to fluid mechanics. Although the dependence between the Hurst exponent and horizon length needs more study over a large variety of datasets which is outside the scope of this work, our initial results on two different, validated DNS datasets show that it can be used as a quantitative \textit{a priori} means of estimating the success of a ROM before we train Recurrent NNs.

\section{Appendix}\label{appendix}

The LSTM NNs have been designed and implemented using the open-source Tensorflow~\cite{abadi2016tensorflow} package, due to its scalable, robust, user-friendly code base. Tensorflow provides a Python API for several of the tensor/matrix manipulation routines it contains, several of which are pre-packaged for deep learning. We used Keras~\cite{chollet2015keras}, a popular Python ``wrapper" for Tensorflow which provides a very concise and simplified API which can considerably automate the NN design process. Keras and Tensorflow, in combination with standard Python libraries for numerical modeling like Numpy and Matplotlib are used to build a self contained program to import datasets and build ROMs. The NN was designed with $1$ hidden layer, making the network intentionally shallow to facilitate comparison with the BiLSTM of the same architecture, and keep computing costs low. Hyperparameter tuning for the number of cells, learning rate adn batch size was performed using a random parameter search. The mean averaged error was chosen as the loss function since it is a regression problem. An ADAM optimizer was found to give superior performance. All the results in this work were obtained by an ensemble of 10 runs  for each case with 75 epochs per run, to account for the stochastic nature of neural networks. The Hurst exponent code was obtained from Ref.~\citen{hurstcode}.

\section*{Supporting Information}
The LSTM code and the associated dataset will be available in the Github repo \url{https://github.com/arvindmohan/LSTM_ROM_Arxiv}.

\nolinenumbers

\bibliography{library}

\bibliographystyle{ieeetr}

\end{document}